\documentclass{aa}
\usepackage{natbib}
\usepackage{graphicx}
\usepackage{times}
\usepackage{amsmath}
\usepackage{txfonts}
\newcommand{\etal}{et al.,~}
\newcommand{\msun}{{\,\rm M}_{\odot}}
\newcommand{\lsun}{{\,\rm L}_{\odot}}

\newcommand{\kms}{\,{\rm km.s}^{-1}}

\newcommand{\nht}{\ifmmode {{\rm NH}_3} \else {NH{\bas 3}} \fi}
\newcommand{\tco}{\ifmmode {^{13}{\rm CO}} \else {$^{13}{\rm CO}$}\fi}
\newcommand{\dco}{\ifmmode {^{12}{\rm CO}} \else {$^{12}{\rm CO}$}\fi}
\newcommand{\cdo}{\ifmmode {{\rm C}^{18}{\rm O}} \else {${\rm C}^{18}{\rm O}$}\fi}

\newcommand{\htco}{\ifmmode {{\rm H}^{13}{\rm CO}^{+} } \else {${\rm H}^{13}
{\rm CO}^{+}$ }\fi}
\newcommand{\hco}{\ifmmode {{\rm H}^{12}{\rm CO}^{+} } \else {${\rm H}^{12}
{\rm CO}^{+}$ }\fi}
\newcommand{\juz}{\ifmmode {{\rm J}=1\rightarrow 0} \else
{J=1$\rightarrow$0}\fi}
\newcommand{\jdu}{\ifmmode {{\rm J}=2\rightarrow 1} \else
{J=2$\rightarrow$1}\fi}
\newcommand{\jtd}{\ifmmode {{\rm J}=3\rightarrow 2} \else
{J=3$\rightarrow$2} \fi}
\newcommand{\jcq}{\ifmmode {{\rm J}=5\!\rightarrow\!4} \else
{${\rm J}=5\!\rightarrow\!4$} \fi}
\newcommand{\as}{\ifmmode {^{\scriptscriptstyle\prime\prime}}
        \else $^{\scriptscriptstyle\prime\prime}$\fi}
\newcommand{\am}{\ifmmode {^{\scriptscriptstyle\prime}}
        \else $^{\scriptscriptstyle\prime}$\fi}
\renewcommand{\hco}{\ifmmode {{\rm HCO}^+} \else {HCO$^+$} \fi}

\newcommand{\tabCylSphere}{
\begin{table}
\caption{Cylindrical vs Spherical description}
 \begin{center}
\begin{tabular}{l c c c}
\hline\hline
 Line
 &  $^{13}$CO(1-0)
 &  $^{13}$CO(2-1)
 &  $^{12}$CO(2-1)\\
\hline
\multicolumn{4}{c}{Cylindrical} \\
\hline \hline
 $R_{out}$ (AU)
 & $    791 \pm      33$
 & $    725 \pm      24$
 & $    868 \pm      15$
 \\
  $T_0$ (K)
 & $      9.1 \pm       0.3$
 & $     14.7 \pm       0.3$
 & $     24.2 \pm       0.2$
 \\
 $q$
 & $     -0.17 \pm       0.09$
 & $      0.18 \pm       0.05$
 & $      0.46 \pm       0.01$
 \\
$V_0$ (km.s$^{-1}$)
 & $      2.12 \pm       0.07$
 & $      2.02 \pm       0.06$
 & $      2.09 \pm       0.06$
 \\
 $v$
 & $      0.5$
 & $      0.5$
 & $      0.5$
 \\
 $dV$ (km.s$^{-1}$)
 & $      0.133 \pm       0.015$
 & $      0.136 \pm       0.015$
 & $      0.147 \pm       0.010$
 \\
 $e_v$
 & $     -0.28 \pm       0.08$
 & $     -0.04 \pm       0.07$
 & $      0.08 \pm       0.04$
 \\
 $\Sigma_m$ (cm$^{-2}$)
 & $    26.9 \pm      1.7\,10^{16}$
 & $    41.4 \pm      6.1\,10^{16}$
 & $    37.4 \pm      9.1\,10^{18}$
 \\
 $p_m$
 & $      3.3 \pm       0.3$
 & $      3.7 \pm       0.4$
 & $      4.5 \pm       0.5$
 \\
  $h_m$ (AU)
 & $     27.1 \pm       4.4$
 & $     27.8 \pm       2.9$
 & $     33.4 \pm       2.3$
 \\
$e_h$
 & $     -1.07 \pm       0.20$
 & $     -1.09 \pm       0.10$
 & $     -0.99 \pm       0.04$
 \\
\hline $\chi^2 $ &  513329.3 &  514557.2 &  136376.5
\\
 \hline \multicolumn{4}{c}{Spherical} \\ \hline \hline
 $R_{out}$
 & $    804 \pm      34$
 & $    737 \pm      20$
 & $    898 \pm      12$
 \\
 $T_0$
 & $      9.3 \pm       0.3$
 & $     15.2 \pm       0.4$
 & $     26.7 \pm       0.2$
 \\
 $q$
 & $     -0.14 \pm       0.09$
 & $      0.20 \pm       0.05$
 & $      0.50 \pm       0.01$
 \\
 $V_0$ (km.s$^{-1}$)
 & $      2.16 \pm       0.07$
 & $      2.07 \pm       0.06$
 & $      2.08 \pm       0.05$
 \\
 $v$
 & $      0.5$
 & $      0.5$
 & $      0.5$
 \\
 $dV$ (km.s$^{-1}$)
 & $      0.135 \pm       0.016$
 & $      0.137 \pm       0.015$
 & $      0.183 \pm       0.012$
 \\
 $e_v$
 & $     -0.26 \pm       0.08$
 & $     -0.05 \pm       0.06$
 & $      0.17 \pm       0.05$
 \\
 $\Sigma_m$ (cm$^{-2}$)
 & $    25.7 \pm      1.8\,10^{16}$
 & $    35.1 \pm      5.3\,10^{16}$
 & $    41.1 \pm      7.4\,10^{17}$
 \\
 $p_m$
 & $      3.2 \pm       0.3$
 & $      3.6 \pm       0.4$
 & $      3.4 \pm       0.4$
 \\
 $h_m$ (AU)
 & $     20.7 \pm       2.2$
 & $     23.4 \pm       2.4$
 & $     19.7 \pm       1.4$
 \\
 $e_h$
 & $     -1.32 \pm       0.18$
 & $     -1.13 \pm       0.12$
 & $     -1.25 \pm       0.06$
 \\
\hline $\chi^2$ &  513329.5 &  514558.9 &  136451.5
 \\
\end{tabular}
\end{center}
{Results for DM\,Tau with ``free'' scale-height and ``free'' line-width (see Sect. \ref{sec:scale} and
\ref{sec:dv} for more details) in the cylindrical and spherical representations. Caution: this table is here
to illustrate the dependence on the geometry of as many parameters as possible. The best parameters to be
applied to DM\,Tau should be taken from Table~\ref{tab:dmtau}.\label{tabcyl-spher}}
\end{table}
}

\newcommand{\TableDM}{
\begin{table*}[!ht]
 \caption{Best parameters for DM\,Tau.}\label{tab:dmtau}
\begin{tabular}{lccccc} 
 \hline \hline
 (1)  & (2) & (3) & (4) & (5) & (6) \\
Lines & \dco~\jdu & \tco~\jdu &\tco~\juz & Mean & \hco~\juz \\
 \hline
 Systemic velocity, $V_\mathrm{LSR}$ (km.s$^{-1}$) &  6.026$\pm 0.002$  & 6.031 $\pm$ 0.003& 6.088 $\pm$ 0.004 & 6.038$\pm 0.002$ & 6.01 $\pm$0.01 \\
 Orientation, PA~($^{\circ}$) &  63.9 $\pm 0.3$ &  65.7  $\pm 0.4 $ & 65.6 $\pm 0.5 $ & 64.7 $\pm$0.2 & $65 \pm 1$\\
 Inclination, $i$~($^{\circ}$) &   -33.5  $\pm 1.0 $ & -35.5  $\pm 1.2 $ & -35.8 $\pm 1.5 $ & -34.7 $\pm 0.7$ & $-33 \pm 2$ \\
 \hline
 \multicolumn{5}{c}{} \\
 \multicolumn{5}{c}{Velocity law:~~~~~~ $V(r) = V_{100} (\frac{r}{100\,\rm{AU}})^{-v}$}\\
& & & \\
 Velocity at 100 AU, $V_{100}$ (km.s$^{-1}$)    & 2.24 $\pm 0.06$ & 2.07 $\pm 0.06$ & 2.15  $\pm 0.08$ & 2.16 $\pm$ 0.04 & $2.10 \pm 0.15$\\
 Stellar mass, M$_*$ ($\msun$) &   0.56 $\pm $0.03  & 0.48 $\pm $0.03  &0.52 $\pm $ 0.04& 0.53$\pm$0.02 & $0.5 \pm 0.1$ \\
 \hline
 Surface Density at 300 AU, (cm$^{-2}$) & $1.4 \pm 0.6\, 10^{17}$ &  $5.3 \pm 1.1\, 10^{15}$ &  $5.9 \pm 0.6\, 10^{15}$ & & $1.1 \pm 0.3\,10^{13}$\\
 Exponent $p$ & $3.8 \pm 0.3$ & $3.4 \pm 0.4$ & $2.8 \pm 0.3$ & & $2.3 \pm 0.4$ \\
 Outer radius $R_\mathrm{out}$, (AU) & $890 \pm 7$ & $720 \pm 19$ & $760 \pm 22$ & & $800 \pm 30$ \\
 Temperature at 100 AU, (K) & $26.0 \pm 0.5$ & $14.5 \pm 0.5$ & $8.0 \pm 0.3$ & & $14 \pm 2$ \\
 Exponent $q$ & $0.49 \pm 0.01$ & $0.12 \pm 0.03$ & $-0.34 \pm 0.07 $ & & $0.40 \pm 0.1$ \\
 dV  (km.s$^{-1}$) & $0.12 \pm 0.01$ & $0.15 \pm 0.01$ & $0.20 \pm 0.01$ & & $0.13 \pm 0.01$ \\
 Scale Height at 100 AU, (AU) & $30 \pm 1.1$ & $28 \pm 4$ & $28 \pm 5$ & & ($19 \pm 8)$\\
 \hline
\end{tabular}\\
Column (1) contains the parameter name. Columns (2,3,4) indicate the parameters derived from \dco~\jdu,
\tco~\jdu\ and \tco~\juz\ respectively. Only \dco\ constrain the size of the inclination. Column (5)
indicates the mean value of the kinematic and geometric parameters. Column (6) indicate the parameters
derived from HCO$^+$~\juz, obtained using a fixed inclination, orientation and stellar mass. For HCO$^+$,
because of the lower resolution, the error on the temperature and surface density include contributions from
the uncertainty on the exponent. $T_m$ is better constrained at 300 AU ($9.0 \pm 0.5$ K), and $\Sigma_m$ at
500 AU ($3.2 \pm 0.5\,10^{12}$ cm$^{-2}$) .
\end{table*}
}

\newcommand{\TableMWC}{
\begin{table*}[!ht]
 \caption{Best parameters for MWC\,480.}\label{tab:mwc480}
\begin{tabular}{lccccc} 
 \hline
 \hline
 (1)  & (2) & (3) & (4) & (5) & (6) \\
Lines & \dco~\jdu & \tco~\jdu &\tco~\juz & \tco & \hco~\juz  \\
 \hline
 Systemic velocity, $V_\mathrm{LSR}$ (km.s$^{-1}$) &  5.076$\pm 0.003$  & 5.16 $\pm$ 0.02& 5.17 $\pm$ 0.02 & 5.084 $\pm$ 0.003  & $5.11 \pm 0.04$\\
 Orientation, PA~($^{\circ}$)  &  58.0 $\pm 0.3$ &  55.8  $\pm 0.6 $ & 57.1 $\pm 0.9 $& 57.8$\pm$0.2 & $ 62 \pm 3$ \\
 Inclination, $i$~($^{\circ}$) &   37.5  $\pm 0.7 $ & 37.6  $\pm 1.6 $ & 33.2 $\pm 3.2 $& 37.0 $\pm$0.6 & $ 34.0 \pm 1.5 $\\
 \hline
 \multicolumn{5}{c}{} \\
 \multicolumn{6}{c}{Velocity law:~~~~~~ $V(r) = V_{100} (\frac{r}{100\,\rm{AU}})^{-v}$}\\
& & & \\
 Velocity at 100 AU,  $V_{100}$ (km.s$^{-1}$)  & 3.95 $\pm 0.06$ & 3.98 $\pm$ 0.14 & 4.45  $\pm 0.33$ & 4.03$\pm$ 0.05 & $3.88 \pm 0.13$ \\
 Stellar mass, M$_*$ ($\msun$)  & 1.76  $\pm $0.06  & 1.65 $\pm $  0.15  & 2.47$\pm $0.23 & 1.83$\pm$ 0.05 & $1.70 \pm 0.12$\\
 \hline
 Surface Density at 300 AU, (cm$^{-2}$) & $3.6 \pm 0.2\, 10^{16}$ &  $3.3 \pm 0.4\, 10^{15}$ &  $5.3 \pm 0.9\, 10^{15}$ &
 $4.5 \pm 0.3\, 10^{15}$ & $3.4 \pm 0.3\,10^{12} $ \\
 Exponent $p$ & $4.7 \pm 0.3$ & $3.0 \pm 0.4$ & $4.5 \pm 0.9$ & $3.9 \pm 0.3 $ & $1.5 \pm 0.2$ \\
 Outer radius $R_\mathrm{out}$, (AU) & $740 \pm 15$ & $450 \pm 15$ & $520 \pm 70$ & $480 \pm 20$ & $ 520 \pm 50 $\\
 Temperature at 100 AU, (K) & $48 \pm 1$ & $28 \pm 2$ & $21 \pm 4$ & $23 \pm 1$ & ($15\pm 6$) \\
 Exponent $q$ & $0.65 \pm 0.02$ & $0.37 \pm 0.08$ & $0.28 \pm 0.09 $ & $0.37 \pm 0.04$ & ($0.6 \pm 0.4$) \\
 dV  (km.s$^{-1}$) & $0.25 \pm 0.01$ & $0.25 \pm 0.02$ & $0.15 \pm 0.03$ & $0.21 \pm 0.02$ & $0.33 \pm 0.07$  \\
 Scale Height at 100 AU, (AU) & $10 \pm 1.1$ & $18 \pm 4$ & $20 \pm 2$ & $19 \pm 2$ & \\
 \hline
\end{tabular}\\
Column (1) contains the parameter name. Columns (2,3,4) indicate the parameters derived from \dco~\jdu,
\tco~\jdu\ and \tco~\juz\ respectively. Column (5) indicates the mean value of the kinematic and geometric
parameters ($V_\mathrm{LSR}$, PA, $i$, $V_{100}$). For the other parameters, Column (5) indicate the results
of a simultaneous fit of the two \tco\ transitions as for LkCa\,15. Column (6) indicates the HCO$^+$ results:
as the temperature is poorly constrained, the surface densities were derived assuming the temperature given
by \tco~\juz{}. The surface density is best determined at 250 AU: $4.3 \pm 0.3\,10^{12}$ cm$^{-2}$.
\end{table*}
}

\newcommand{\TableLKCA}{
\begin{table*}[!ht]
 \caption{Best parameters for LkCa\,15}\label{tab:lkca}
\begin{tabular}{lcccc} 
 \hline \hline
 (1)  & (2) & (3) & (4) & (5)  \\
Lines & \dco~\jdu & \tco & Mean & \hco~\juz  \\
 \hline
 Systemic velocity, $V_\mathrm{LSR}$ (km.s$^{-1}$)  & 6.29$\pm 0.01$  & 6.33 $\pm$ 0.02 & 6.30 $\pm$0.01 & $6.26 \pm 0.04$ \\
 Orientation, PA~($^{\circ}$) &   150.9 $\pm 0.4$ &  150.4  $\pm 1.0 $ & 150.7 $\pm$0.4 & $150 \pm 2$\\
 Inclination, $i$~($^{\circ}$)  &  51.8  $\pm 0.8 $ & 49.6  $\pm 1.0 $ & 51.5$\pm$0.7 & $51.5 \pm 1.3$ or $46 \pm 3$\\
 \hline
 \multicolumn{4}{c}{} \\
 \multicolumn{4}{c}{Velocity law:~~~~~~ $V(r) = V_{100} (\frac{r}{100\,\rm{AU}})^{-v}$}\\
& & & \\
 Velocity at 100 AU, $V_{100}$ (km.s$^{-1}$)   &  3.07 $\pm 0.06$ & 2.97 $\pm 0.06$ & 3.00 $\pm$0.05 & $2.91\pm 0.05$ or $3.11 \pm 0.15$ \\
 Stellar mass, M$_*$ ($\msun$)  &  1.06 $\pm$ 0.04 & 0.99 $\pm $ 0.05 & 1.01 $\pm$ 0.03 & \\
 \hline
 \multicolumn{4}{c}{} \\
 Surface Density at 300 AU, (cm$^{-2}$) & $2.9 \pm 0.1\, 10^{16}$ &  $2.6 \pm 0.2\, 10^{15}$ & & $8.0\pm 0.5\,10^{12}$ \\ 
 Exponent $p$ & $4.4 \pm 0.3$ & $1.50 \pm 0.15$  & & $2.5 \pm 0.3$ \\ 
 Outer radius $R_\mathrm{out}$, (AU) & $905 \pm 40$ & $550 \pm 20$ & & $ 660 \pm 60$\\
 Temperature at 100 AU, (K) & $22.1 \pm 0.5$ & $21.2 \pm 1.7$ & & [19] \\ 
 Exponent $q$ & $0.37 \pm 0.02$ & $0.40  \pm 0.10$ & & [0.38] \\ 
 dV  (km.s$^{-1}$) & $0.19 \pm 0.01$ & $0.29 \pm 0.03$ & & $0.23 \pm 0.04$ \\
 Scale Height at 100 AU, (AU) & $19 \pm 1$ & $17 \pm 3$ & & ($ \pm $)\\
 \hline
\end{tabular}\\
Column (1) contains the parameter name. Columns (2,3) indicate the parameters derived from \dco~\jdu, and a
combined fit of the \tco~\jdu\ and \tco~\juz\ respectively. Column (4) indicates the mean value of the
kinematic and geometric parameters. Column (5) indicates the HCO$^+$ results: we used a fixed temperature law
compatible with the CO isotopologue results (see text).
\end{table*}
}

\newcommand{\TableROUT}{
\begin{table}[!ht]
 \caption{HCO$^+$ and CO Outer Radius}\label{tab:out}
\begin{tabular}{lccc}
 \hline
 \hline
 Source & $R_{out}$(\dco) & $R_{out}$(\tco) & HCO$^+$\\
 & (AU) & (AU) & (AU) \\
 \hline
 DM\,Tau  & $890\pm 7$ & $740 \pm 15$ & $800 \pm 80$ \\
 LkCa\,15 & $905\pm 40$ & $550 \pm 20$ & $660 \pm 60$ \\
 MWC\,480 & $740\pm 15$ & $480 \pm 20$ & $520 \pm 50$\\
\hline
 \end{tabular}
\end{table}
}

\newcommand{\figMAPS}{
\begin{figure*}[ht]
  \centering
  \resizebox{14.0cm}{!}{
  \includegraphics[angle=0.0,width=12cm]{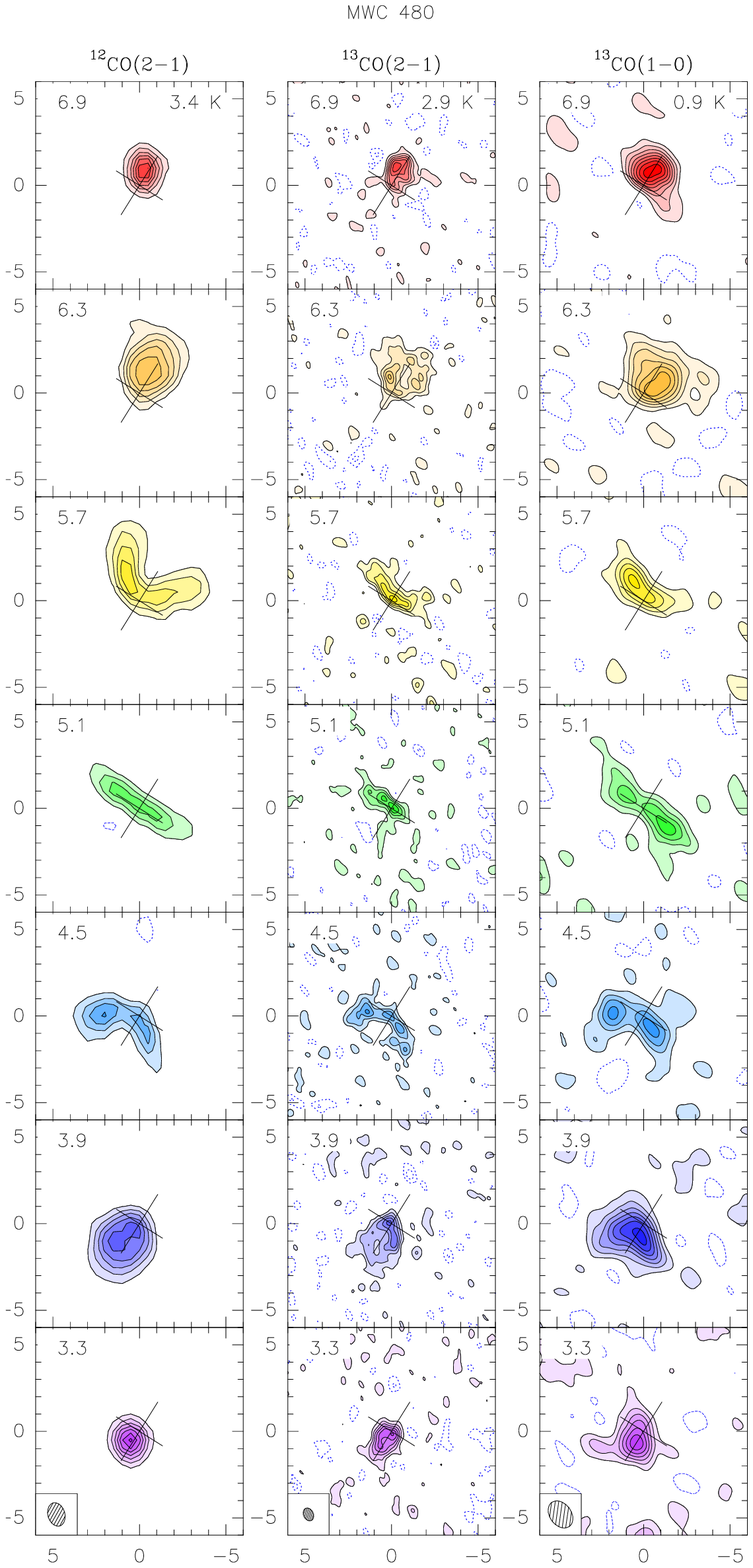} 
  \includegraphics[width=12cm,angle=0]{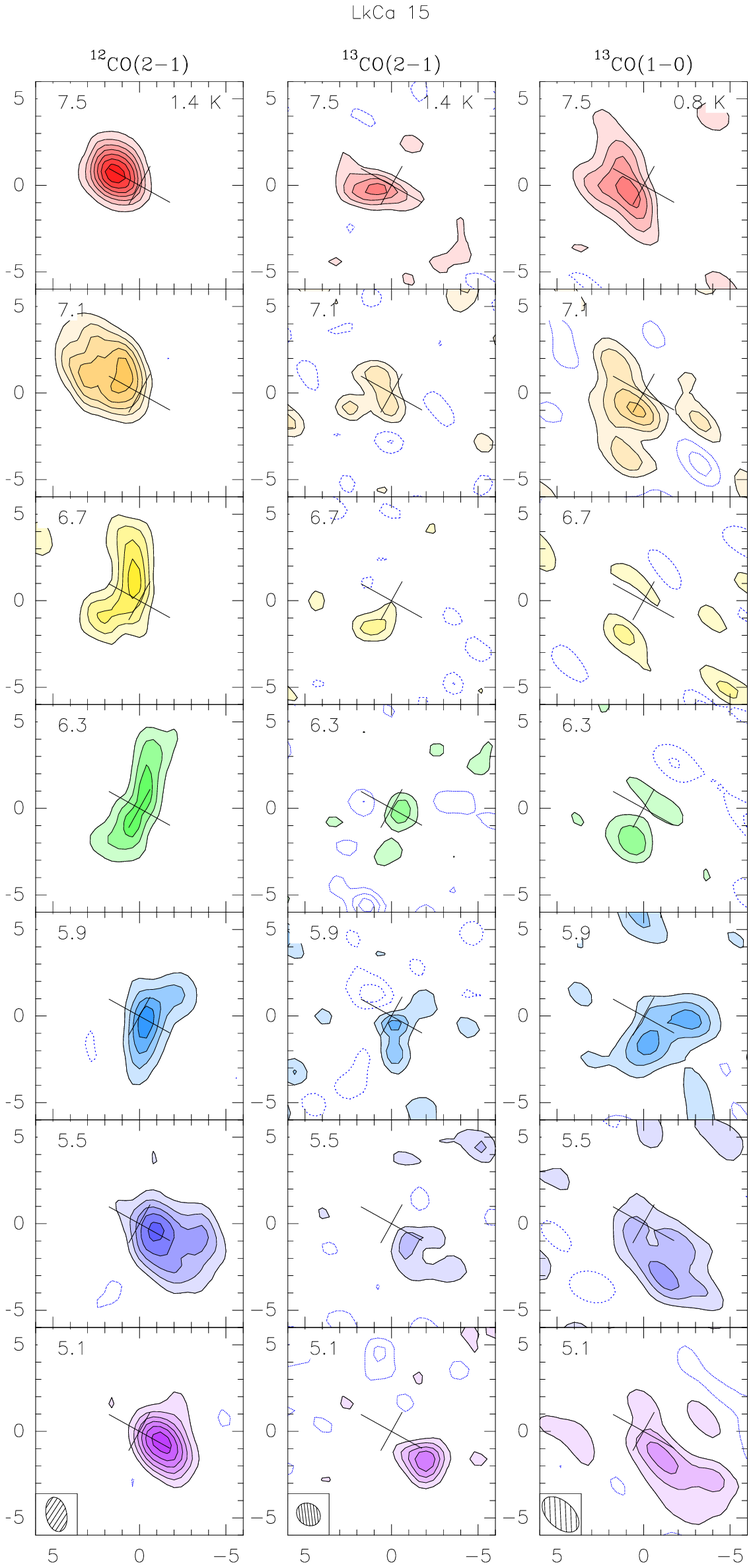}} 
  \caption{Channel maps of
 the \dco~and~\tco~ line emission towards MWC\,480 and LkCa\,15.  The
 synthesized beam is indicated in each panel. The cross indicates the
 direction of the major and minor axis of each disk, as respectively
 derived from the analysis of the line data.  The velocity of the
 channel is indicated in the upper left corner of each panel (LSR
 velocity in km.s$^{-1}$).  A 200 m taper has been applied on the
 \tco~\jdu~ data of LkCa\,15 which where obtained at higher angular
 resolution. {\bf Left: MWC\,480} Left: \dco\,\jdu\ line, spatial
 resolution $1.38 \times 0.99''$ at PA~$16^\circ$, contour spacing 200
 mJy/beam, or 3.4~K, or 3.6 $\sigma$.  Middle: \tco\,\jdu\ line,
 spatial resolution $0.77 \times 0.57$ at PA~$26^\circ$, contour
 spacing 50 mJy/beam, or 2.9~K, or 1.9 $\sigma$.  Right: \tco\,\juz\
 line, spatial resolution $1.72 \times 1.24''$ at PA~$37^\circ$,
 contour spacing 20 mJy/beam, or 0.9~K, or 1.8 $\sigma$. Negative
 contours are dashed, and the zero contour is omitted.  {\bf Right:
 LkCa\,15} Left: \dco\,\jdu\ line, spatial resolution $2.01 \times
 1.20''$ at PA~$12^\circ$, contour spacing 150 mJy/beam, or 1.4~K, or
 2.5 $\sigma$.  Middle: \tco\,\jdu\ line, spatial resolution $1.46
 \times 1.24$ at PA~$52^\circ$, contour spacing 100 mJy/beam, or
 1.4~K, or 1.6 $\sigma$.  Right: \tco\,\juz\ line, spatial resolution
 $2.55 \times 1.51''$ at PA~$47^\circ$, contour spacing 30 mJy/beam,
 or 0.8~K, or 1.8 $\sigma$. Negative contours are dashed, and the zero
 contour is omitted.  }
\label{fig:maps}
\end{figure*}
}

\newcommand{\figHCO}{
\begin{figure*}[!t]
  \centering
  \includegraphics[width=7.0cm,angle=270]{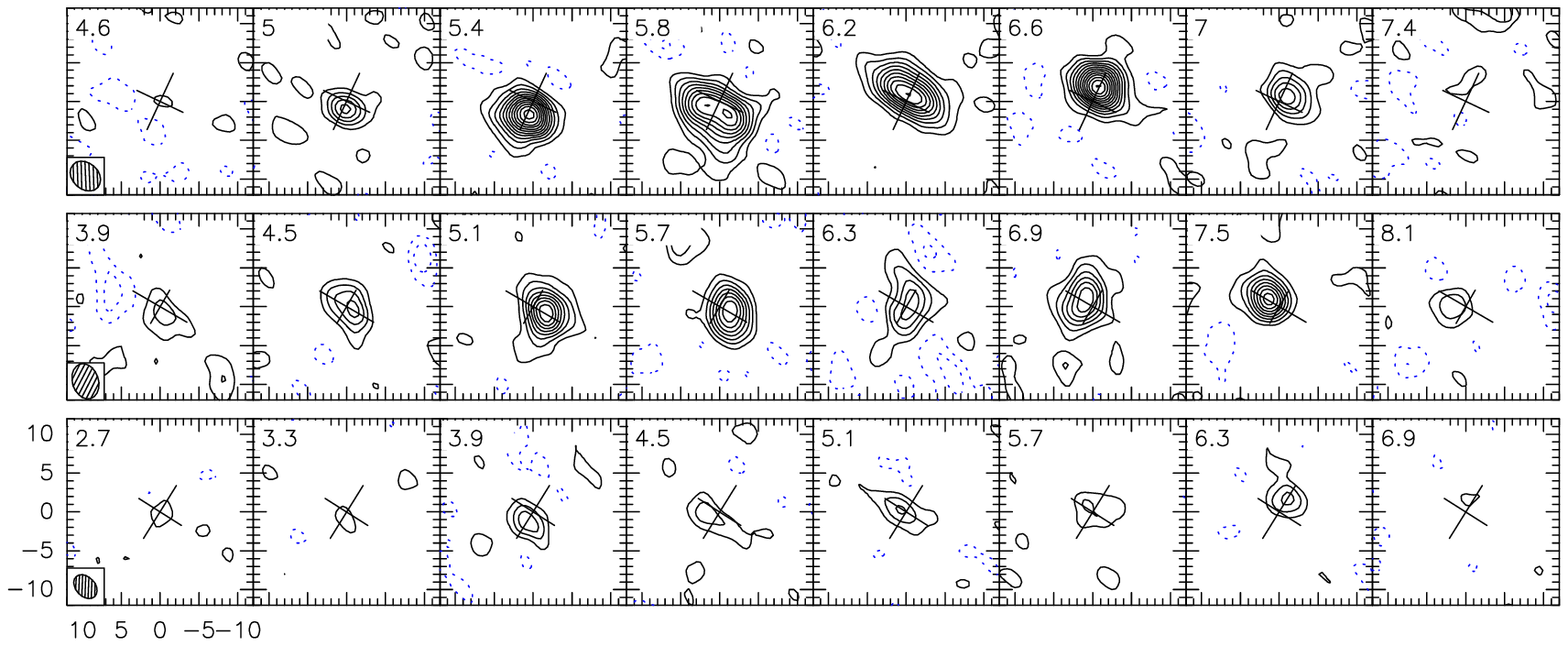} 
  \caption{Channel maps of the HCO$^+$~\juz{} line emission towards
 DM\,Tau, LkCa\,15 and MWC\,480. The cross indicates the direction of
 the major and minor axis of each disk. The velocity of the channel is
 indicated in the upper left corner of each panel (LSR velocity in
 km.s$^{-1}$).  For displaying purpose, the spectral resolution is 0.6
 $\kms$ for MWC\,480 and LkCa\,15, and 0.4 $\kms$ for DM\,Tau. All contour steps
 are 20 mJy/beam ($\sim 2 \sigma$), negative contours are dashed and the
 zero contour is omitted.
 Top: DM\,Tau, spatial resolution is $4.3 \times
 3.3''$ at PA~$47^\circ$, contour step 0.22 K.
 Middle: LkCa\,15, spatial
 resolution is $4.6 \times 3.4''$ at PA~$19^\circ$, contour step 0.25 K. Bottom:
 MWC\,480, spatial resolution is $3.5 \times 2.5''$ at PA~$42^\circ$,
 contour step 0.36 K.  }
  \label{fig:hco}
\end{figure*}
}

\newcommand{\convention}{
\begin{figure}[!ht]
\includegraphics[width=9cm]{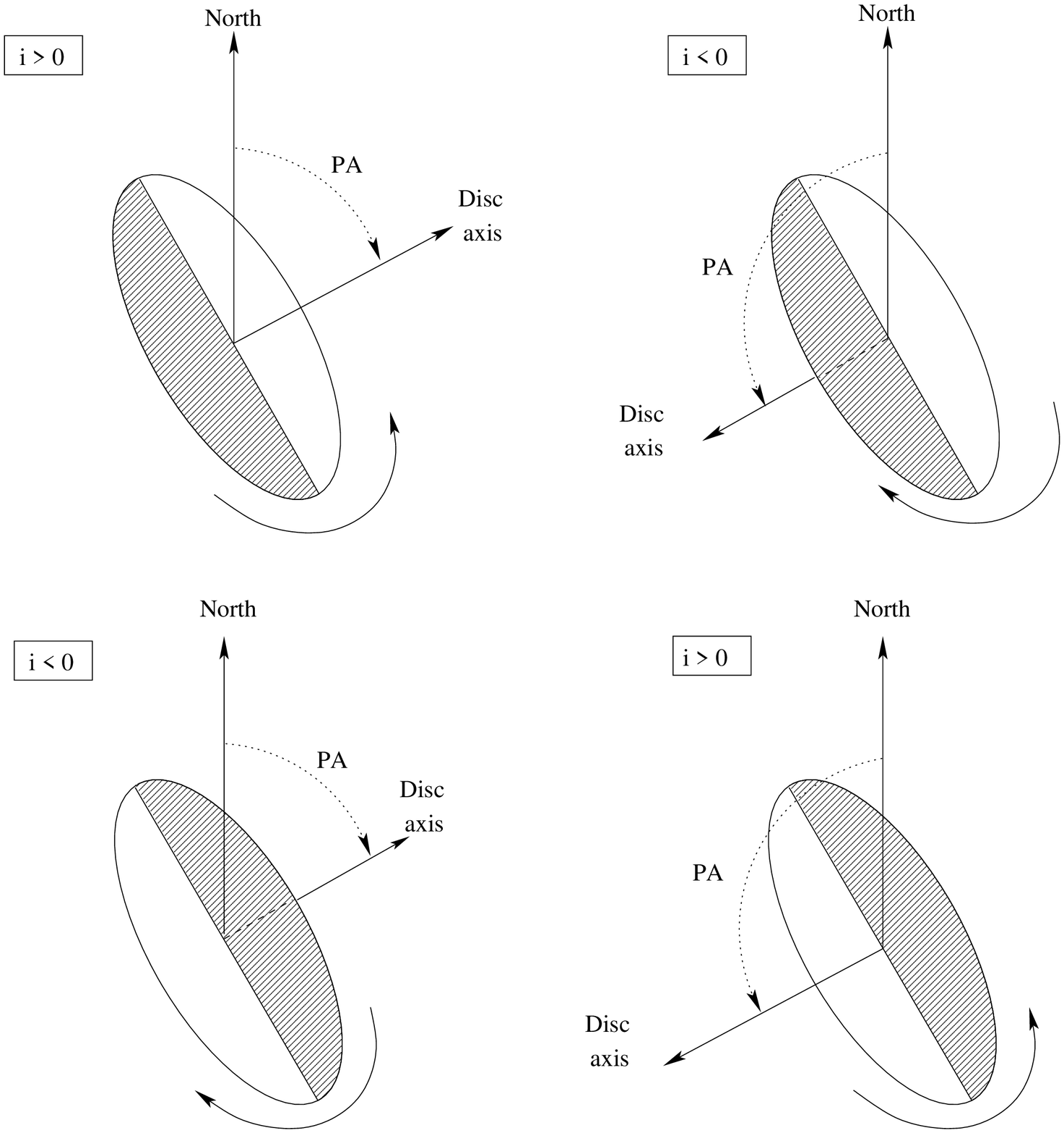} 
\caption{Geometrical convention: four disk configurations yielding the same aspect ratio when projected are
presented. The shaded area correspond to the part of the disk that is closer to us than the plane of the sky
containing the disk axis. The curved arrow represents the sense of rotation (this defines the disk axis). The
projected velocities allow to disentangle between one of the two columns (for the two cases presented on the
left, the red-shifted part would be the lower part, and the contrary for the two cases on the right). It is
possible to determine the sign of the inclination either with the asymmetry at the systemic velocity if
sufficient signal to noise is available, or by other means, e.g. scattered images in the optical and near-IR.
PA is classically designed to be positive East from North.} \label{fig:geo}
\end{figure}
}

\newcommand{\radiusDMTAU}{
\begin{figure*}
\includegraphics[width=4.0cm,angle=270]{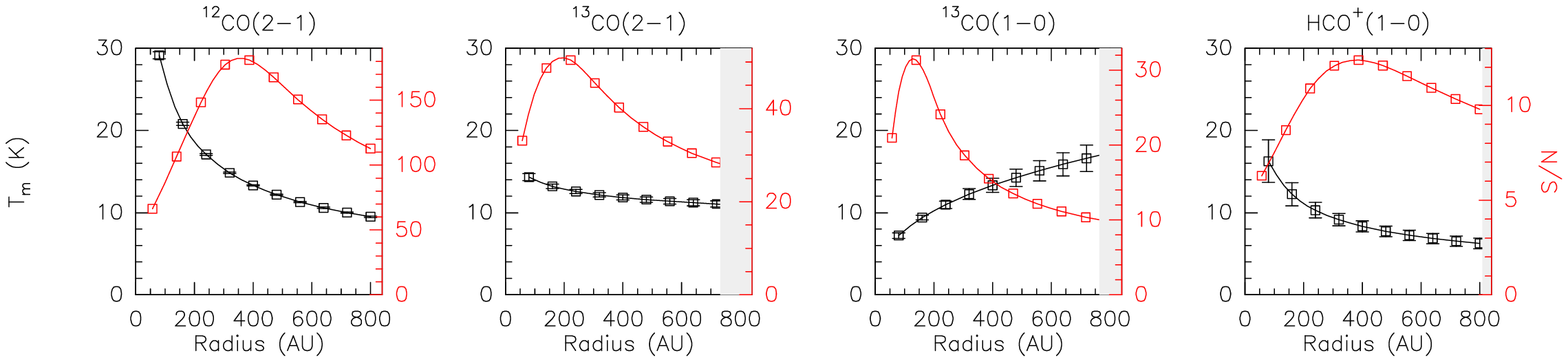}\\ 
\includegraphics[width=4.0cm,angle=270]{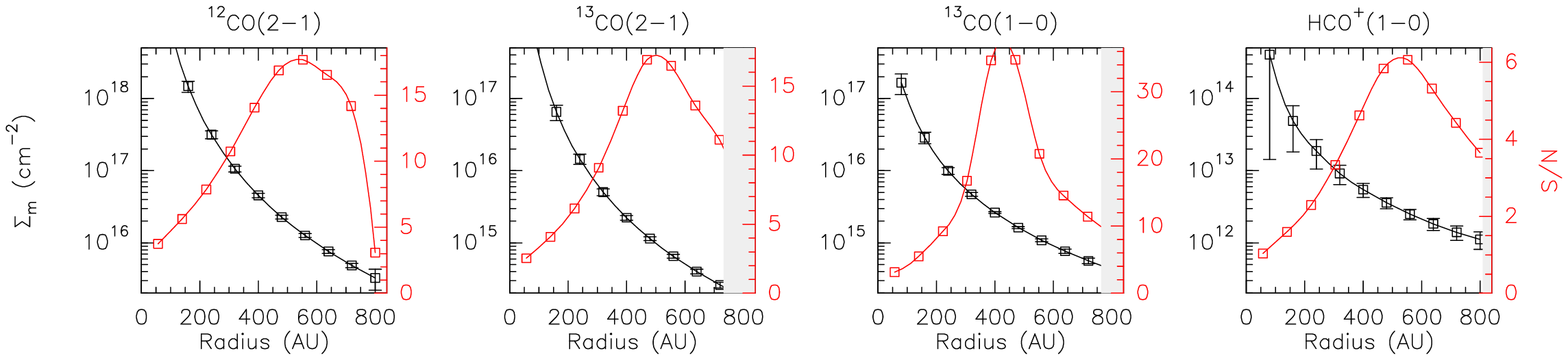}\\ 
\includegraphics[width=4.0cm,angle=270]{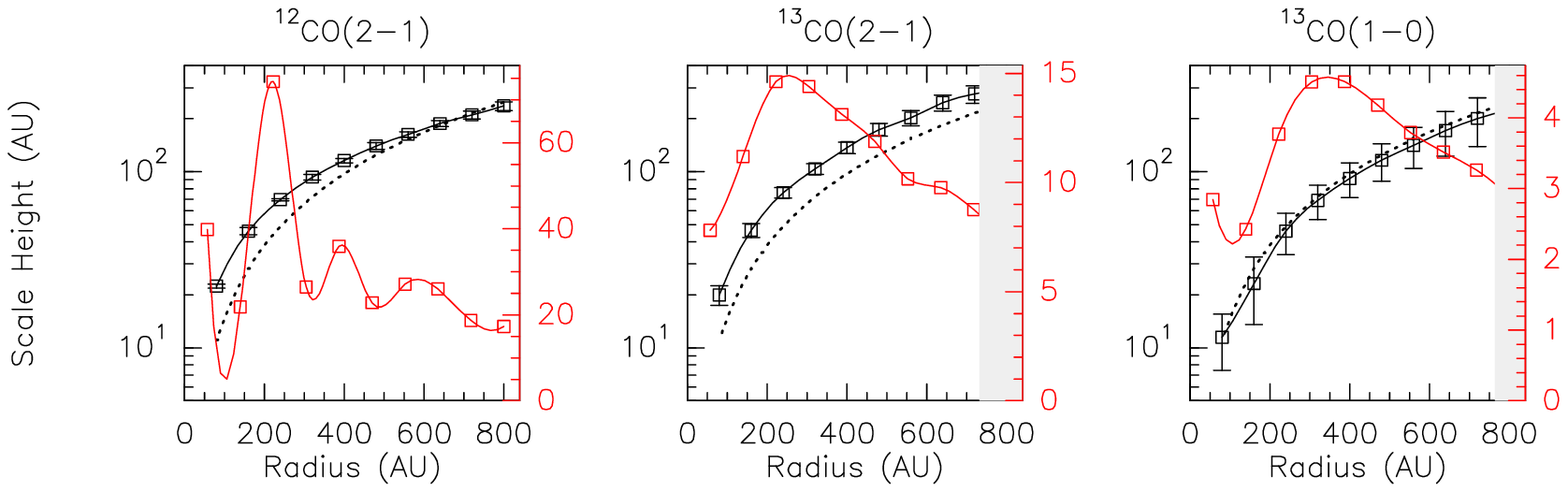} 
{\parbox[t][2.0cm][b]{6.0cm}{\Large\bf ~~~~DM~Tau}} \caption{ From left to right \dco~\jdu, \tco~\jdu,
\tco~\juz~ and \hco~\juz~ in DM\,Tau. {\bf Top:} Temperature $T_\mathrm{m}$ (black curve, left axis) and
signal to noise on the temperature $T_\mathrm{m} / \delta T_\mathrm{m}$ (grey curve, right axis) as a
function of reference radius $R_T$.{\bf Middle} Surface density $\Sigma_m$ (left axis) and signal to noise on
the surface density $\Sigma_m/\delta \Sigma_m$ (right axis) as a function of reference radius $R_\Sigma$.
{\bf Bottom} Scale height $h$ (left axis) and signal to noise on the scale height $h/\delta h$ (right axis)
as a function of reference radius $R_h$. \label{fig:raddm} }
\end{figure*}
}
\newcommand{\radiusLKCA}{
\begin{figure*}
\includegraphics[width=4.0cm,angle=270]{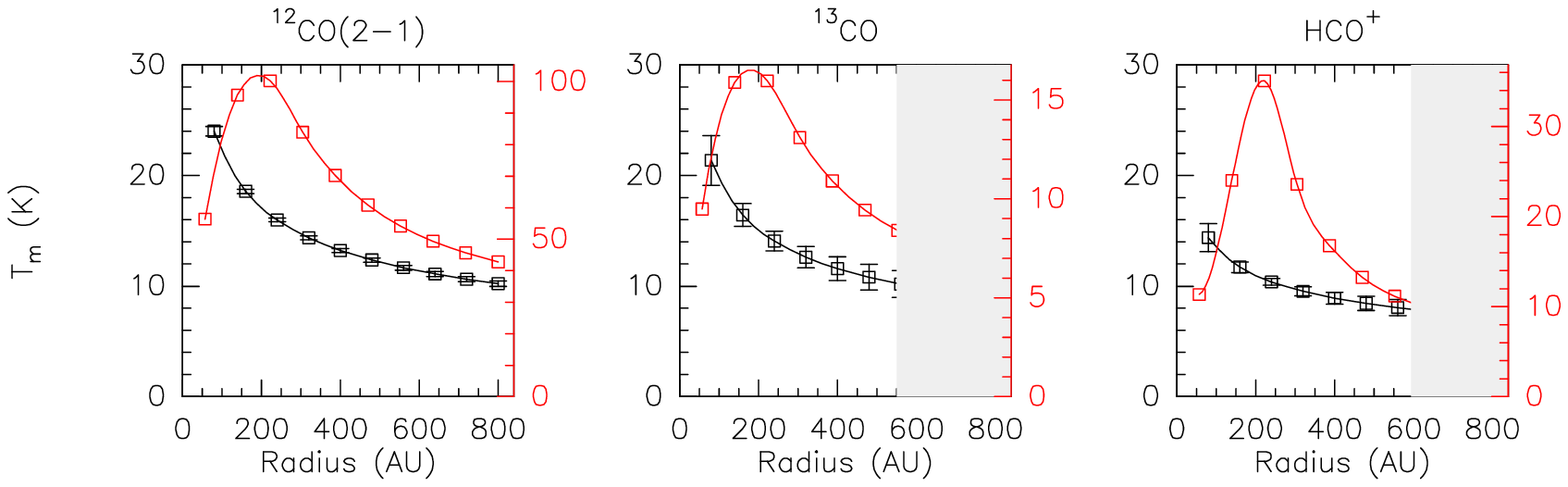} \\ 
\includegraphics[width=4.0cm,angle=270]{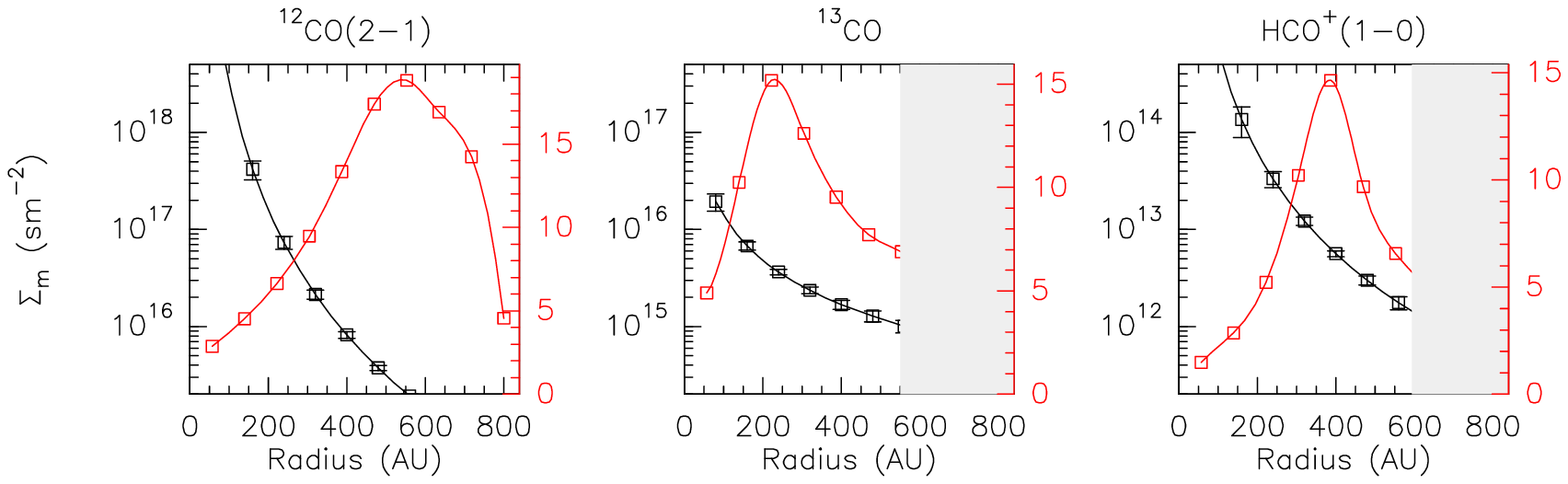} \\ 
\includegraphics[width=4.0cm,angle=270]{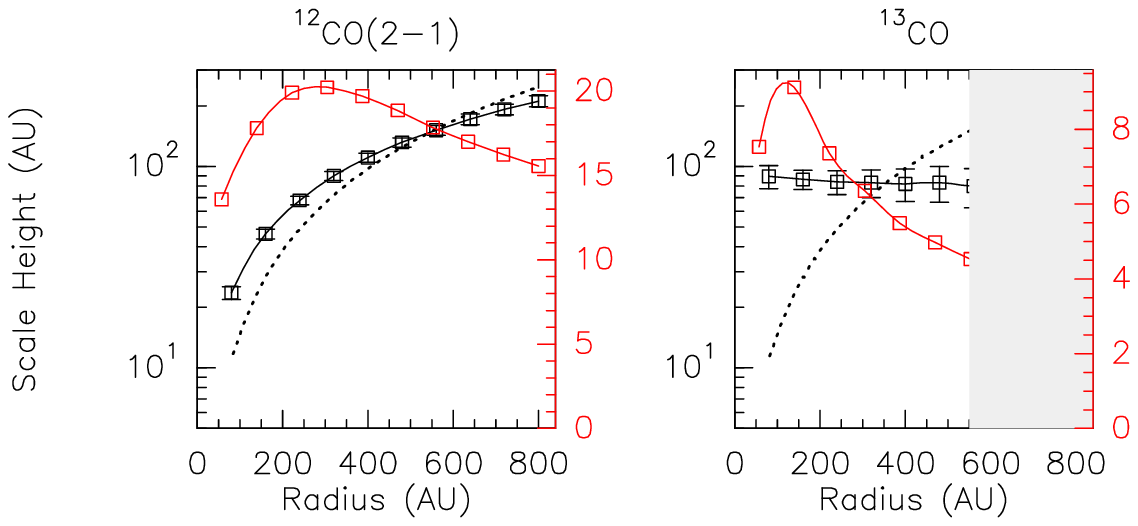} 
{\parbox[t][2.0cm][b]{6.0cm}{\Large\bf ~~~~LkCa\,15}} \caption{ From left to right \dco, \tco~ and \hco~\juz~
in LkCa\,15. {\bf Top:} Temperature $T_\mathrm{m}$ (black curve, left axis) and signal to noise on the
temperature $T_\mathrm{m} / \delta T_\mathrm{m}$ (grey curve, right axis) as a function of reference radius
$R_T$.{\bf Middle} Surface density $\Sigma_m$ (left axis) and signal to noise on the surface density
$\Sigma_m/\delta \Sigma_m$ (right axis) as a function of reference radius $R_\Sigma$. {\bf Bottom} Scale
height $h$ (left axis) and signal to noise on the scale height $h/\delta h$ (right axis) as a function of
reference radius $R_h$. For \hco~ the temperature and surface density were adjusted separately: the curves
should be used as an indicator of the region over which these values are actually constrained, but not as a
quantitative measure in terms of S/N, as the coupling between temperature and surface density is ignored.
\label{fig:radlk} }
\end{figure*}
}
\newcommand{\radiusMWC}{
\begin{figure*}
\includegraphics[width=4.0cm,angle=270]{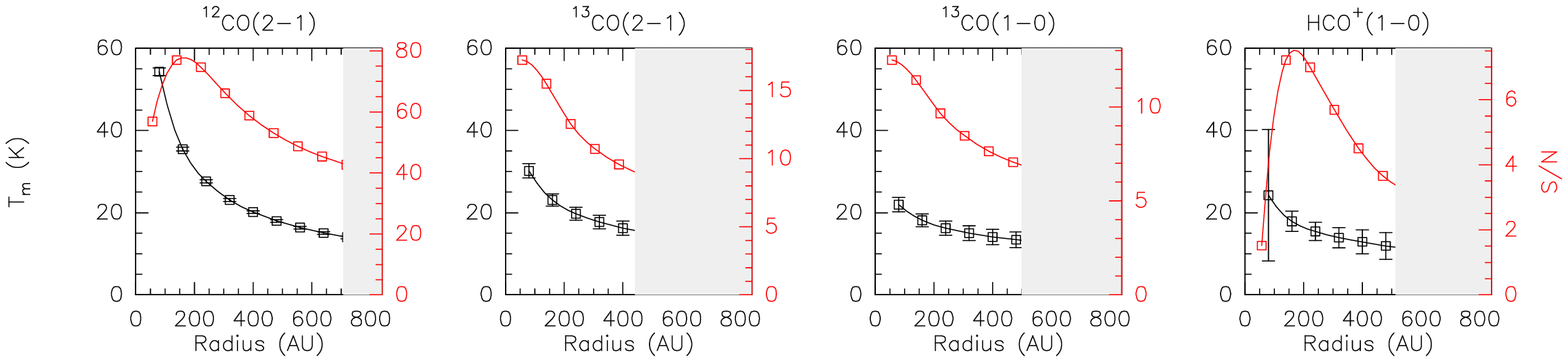} \\ 
\includegraphics[width=4.0cm,angle=270]{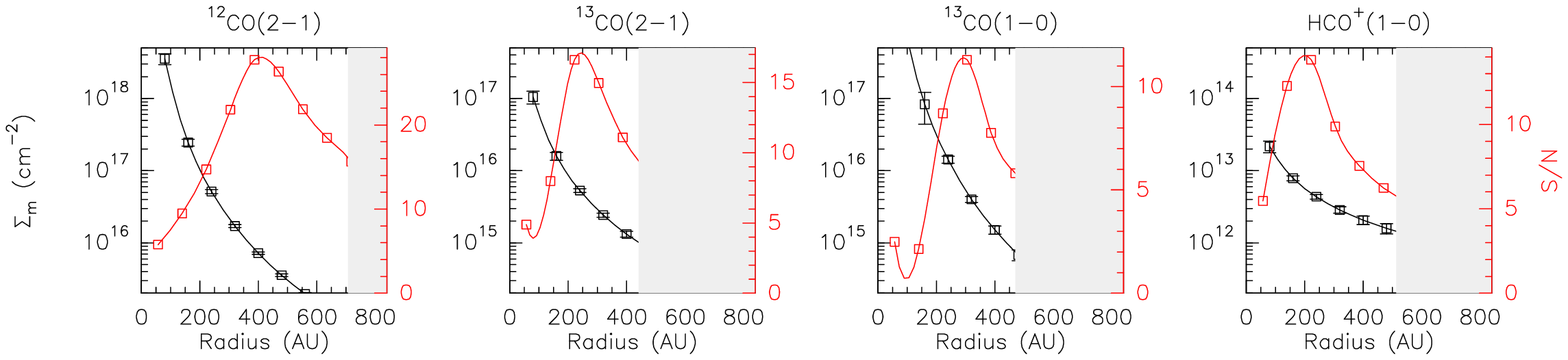} \\ 
\includegraphics[width=4.0cm,angle=270]{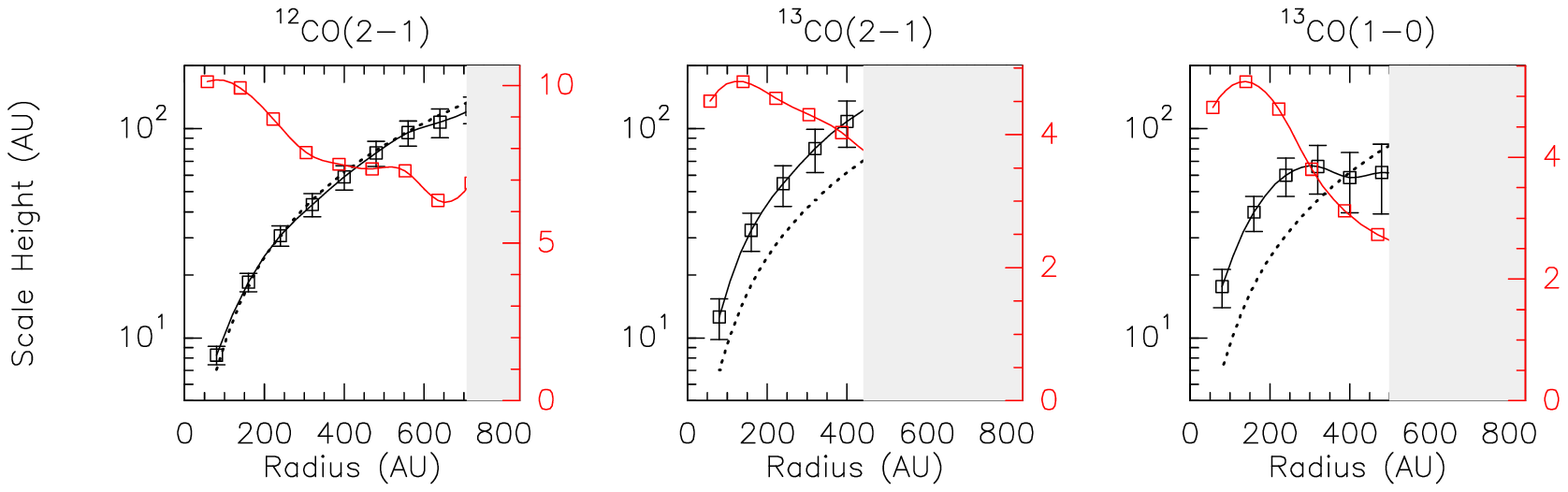} 
{\parbox[t][2.0cm][b]{6.0cm}{\Large\bf ~~~~MWC~480}} \caption{ From left to right \dco~\jdu, \tco~\jdu,
\tco~\juz~ and \hco~\juz~in MWC\,480. {\bf Top:} Temperature $T_\mathrm{m}$ (black curve, left axis) and
signal to noise on the temperature $ T_\mathrm{m} / \delta T_\mathrm{m}$ (grey curve, right axis) as a
function of reference radius $R_T$.{\bf Middle} Surface density $\Sigma_m$ (left axis) and signal to noise on
the surface density $\Sigma_m/\delta\Sigma_m$ (right axis) as a function of reference radius $R_\Sigma$. {\bf
Bottom} Scale height $h$ (left axis) and signal to noise on the scale height $h/\delta h$ (right axis) as a
function of reference radius $R_h$. As for LkCa~15, temperature and surface density are adjusted separately
for HCO$^+$. \label{fig:radmw} }
\end{figure*}
}
\newcommand{\geom}{
\begin{figure*}[htb]
\centering
\includegraphics[angle=270]{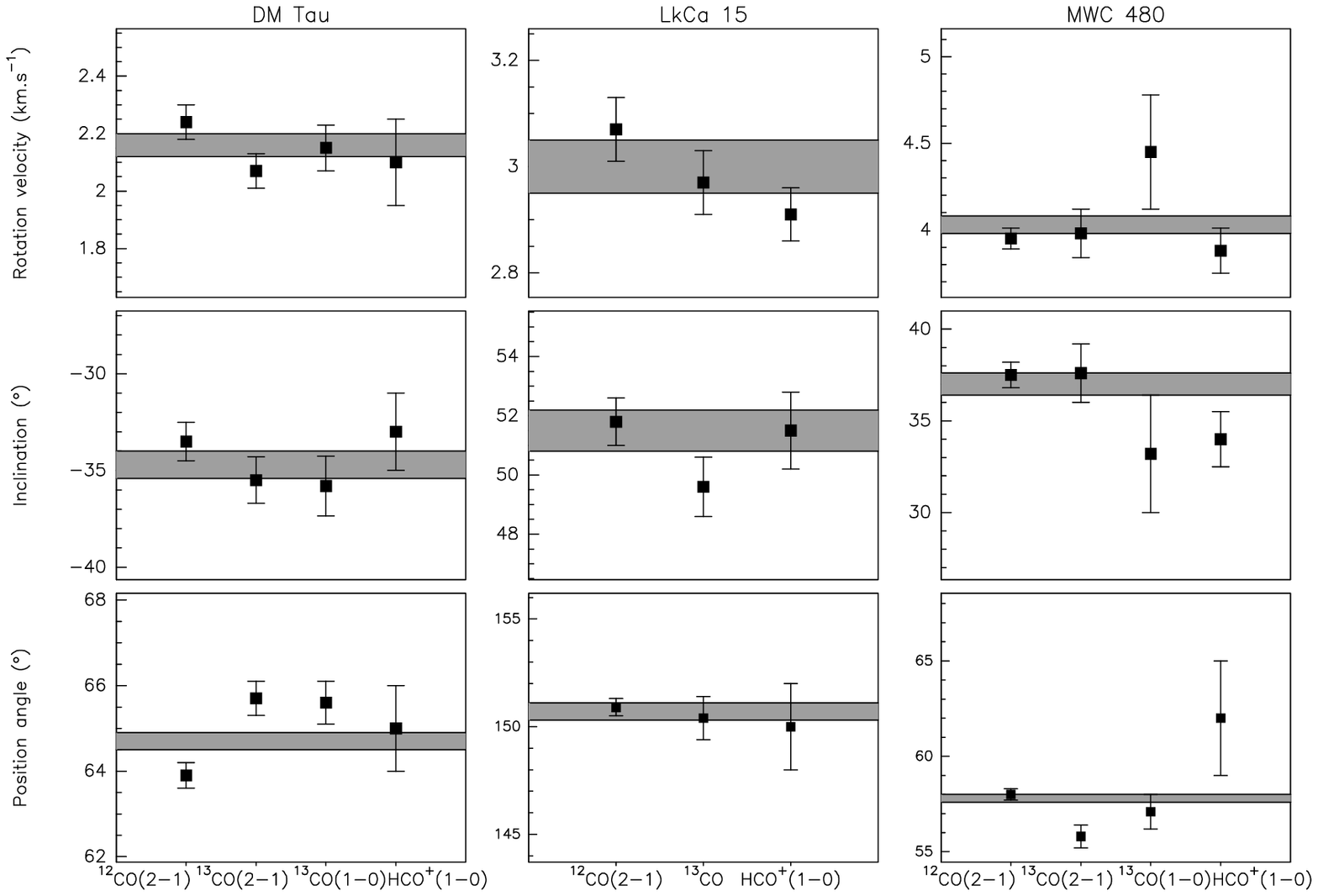} 
\caption{From left to right: kinetic and geometric parameters derived for DM\,Tau, LkCa\,15 and MWC\,480.
From top to bottom: rotation velocity at 100~AU, inclination and position angle. For DM\,Tau and MWC\,480,
the error bars represent (from left to right) the value of the parameter measured from the \dco~\jdu,
\tco~\jdu, \tco~\juz~ and \hco~\juz~ lines. For LkCa\,15 (from left to right), the values are derived from
the \dco~\jdu~ line, from the simultaneous fit of the \tco~\jdu~ and the \tco~\juz~ lines, and from the
\hco~\juz~ line. In all three sources, the grey area represent $\pm 1 \sigma$ around the mean of the values
of the CO isotopes. \label{fig:para} }
\end{figure*}
}

\newcommand{\FigMASS}{
\begin{figure*}[ht]
  \resizebox{18.0cm}{!}{
\includegraphics[angle=270]{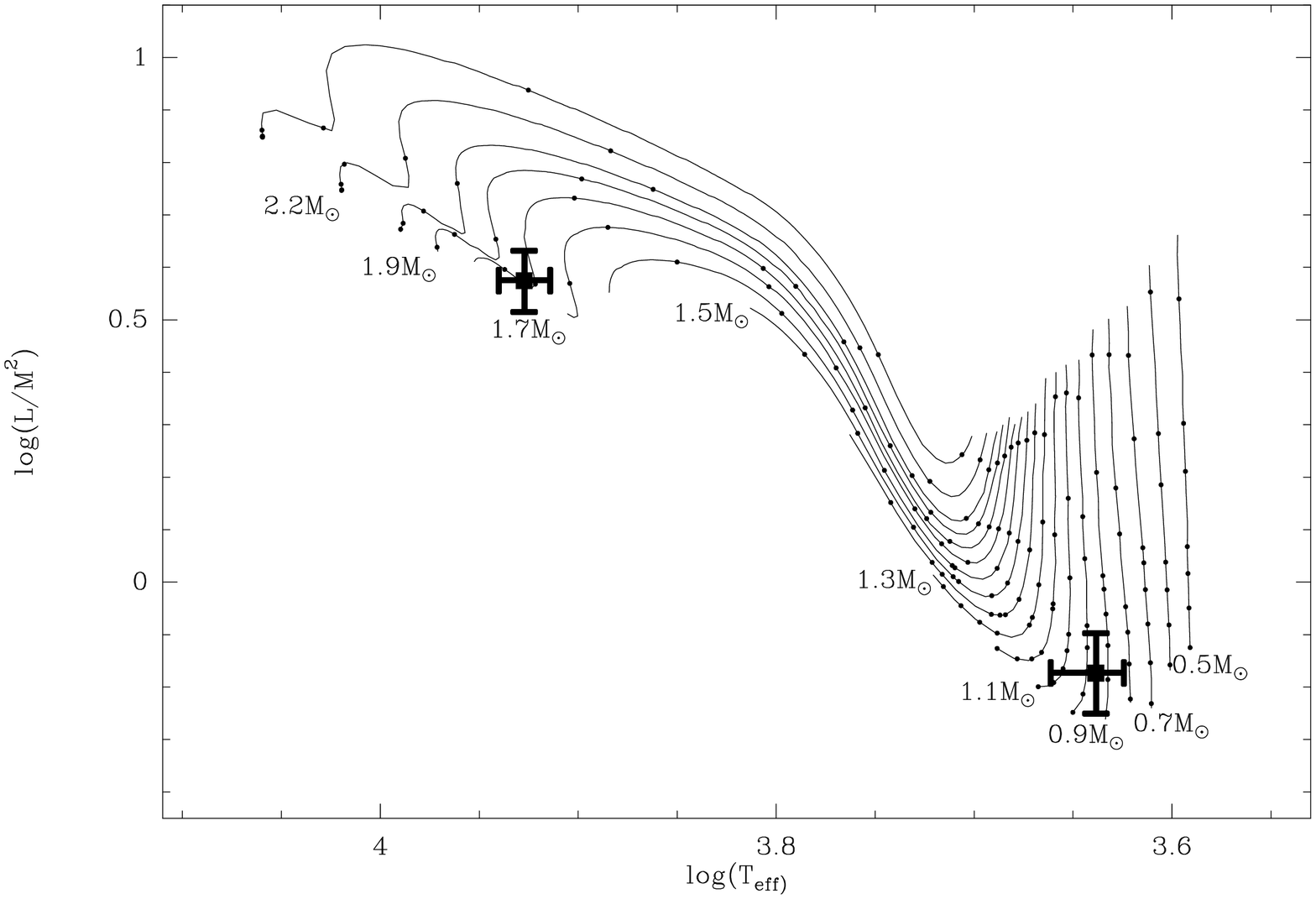} 
\includegraphics[angle=270]{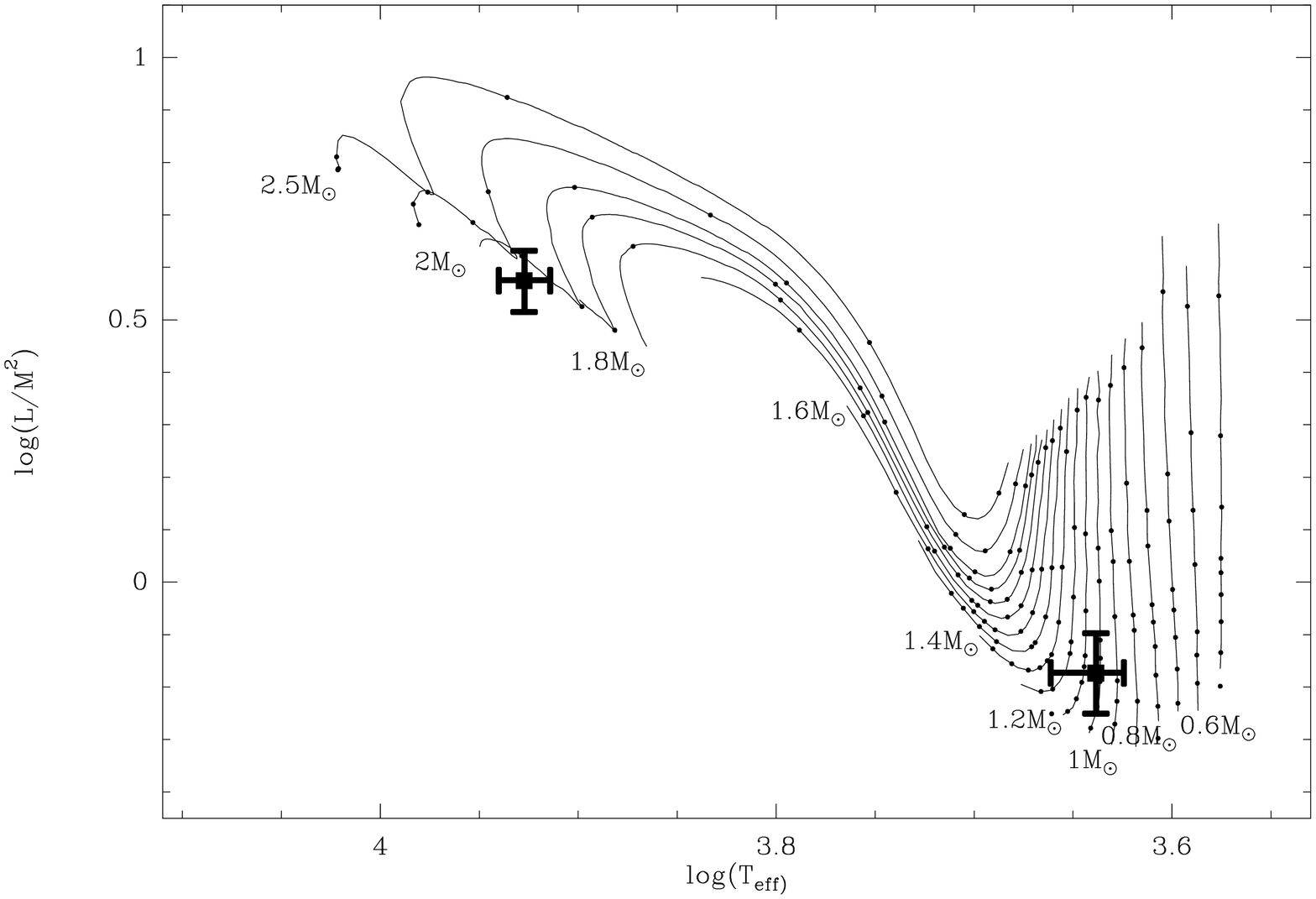}} 
  \caption{Summary of the dynamical mass measurement for the sources
of this sample. As in \citet{Simon_etal2000}, the results are shown in a distance corrected HR diagram
($L/M^2$ versus T$_{eff}$). New results have been obtained for MWC\,480 and LkCa\,15. They are all compatible
with the previous ones. In the case of MWC\,480, the improvement in the mass determination is due to a better
fit of the continuum, as explained in Sect.\ref{sec:fitcont}. {\bf Left:} Tracks for a stellar metallicity
$Z=0.01$. {\bf Right:} Tracks for a stellar metallicity $Z=0.02$. The models have been computed by
\citet{Siess_etal2000}. Dots starts at 1 Myr and are spaced by 1 Myr. The tracks range from 0.5 to 2.0 by 0.1
$\msun$, with the 2.2 and 2.5 $\msun$ tracks added. For T$_{eff}$, the  error bars are assumed to be +/- one
sub spectral type class.}
  \label{fig:hrd-mass}
\end{figure*}
}

\newcommand{\FigTEMP}{
\begin{figure}
  \resizebox{\columnwidth}{!}{\includegraphics[angle=270]{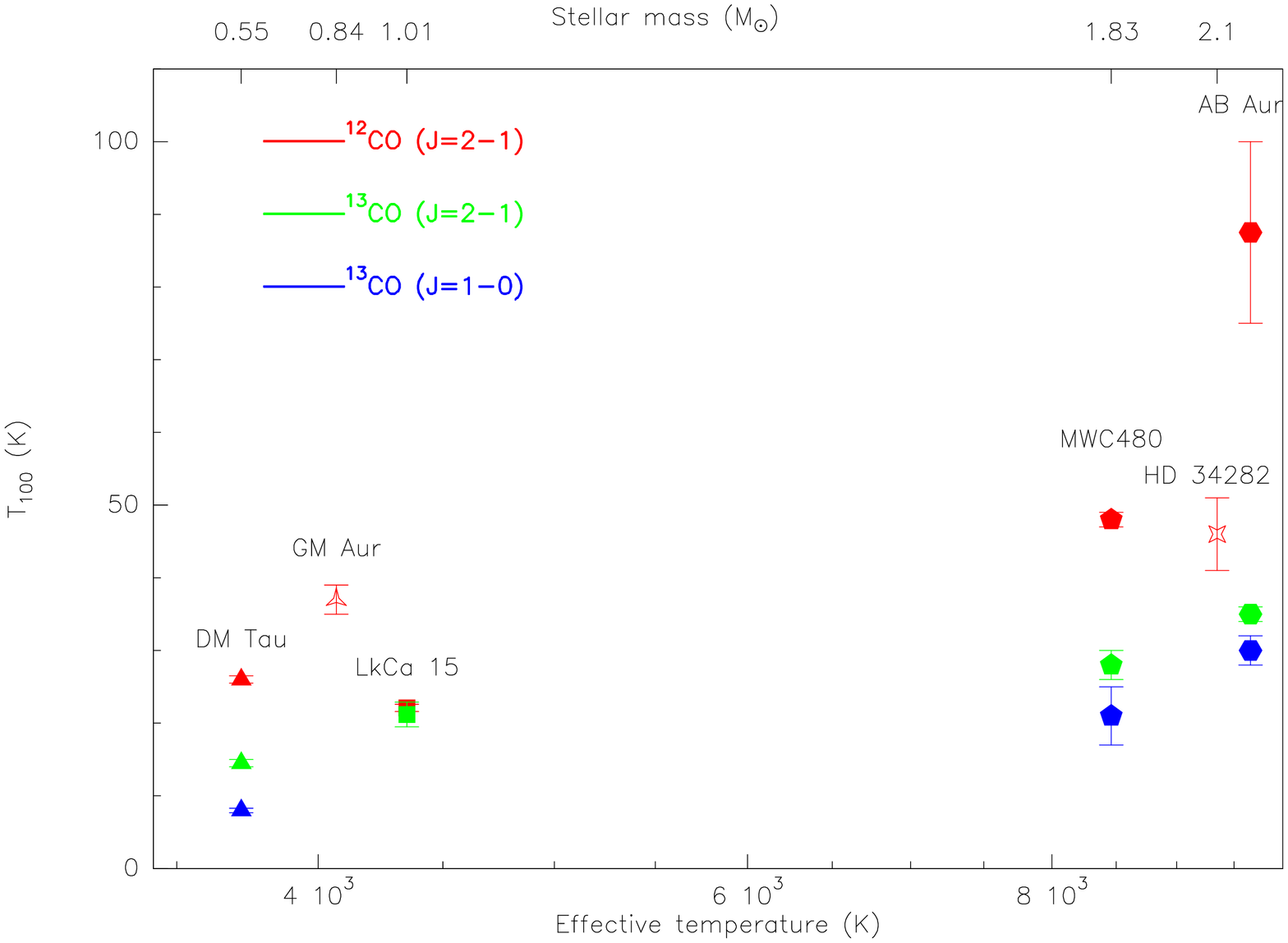}} 
\caption{Temperatures derived from the CO isotopes versus effective temperature of the central star. From
left to right, the sources are DM\,Tau (filled triangles), GM Aur (empty stars), LkCa\,15 (filled squares),
MWC\,480 (filled pentagons), HD 34282 (empty stars) and AB\,Aur (filled hexagons). From top to bottom,
\dco~\jdu, \tco~\jdu, \tco~\juz~temperatures.}
  \label{fig:temp}
\end{figure}
}

\begin{document}
\title{Probing the structure of protoplanetary disks:
a comparative study of DM~Tau, LkCa~15 and MWC~480}
\author{Vincent Pi\'etu\inst{1,2}, Anne Dutrey\inst{1} and
St\'ephane Guilloteau \inst{1}}
\offprints{V.Pi\'etu, \email{pietu@iram.fr}}
\institute{Université Bordeaux 1 ; CNRS ; OASU ; UMR 5804, BP 89, 2 rue de l'Observatoire, F-33270 Floirac,
France \and Institut de Radio-Astronomie Millim\'etrique, 300 rue de la Piscine, Domaine Universitaire
F-38406 Saint Martin d'H\`eres, France}
\date{Received 11-Oct-2006, Accepted 08-Jan-2007}
\authorrunning{Pi\'etu \etal}
\titlerunning{Vertical structure of protoplanetary disks surrounding HAe and
TTauri stars}

  \abstract
{The physical structure of proto-planetary disks is not yet well constrained by current observations.
Millimeter interferometry is an essential tool to investigate young disks.}
{We study from an observational perspective the vertical and radial temperature distribution in a few
well-known disks.  The surface density distribution of CO and HCO$^+$ and the scale-height are also
investigated.}
{We report CO observations at sub-arcsecond resolution with the IRAM array of the disks surrounding MWC~480,
LkCa~15 and DM~Tau, and simultaneous measurements of HCO$^+$~\juz. To derive the disk properties, we fit to
the data a standard disk model in which all parameters are power laws of the distance to the star. Possible
biases associated to the method are detailed and explained. We compare the properties of the observed disks
with similar objects.} 
{We find evidence for vertical temperature gradient in the disks of MWC\,480 and DM\,Tau, as in AB\,Aur, but
not in LkCa\,15. The disks temperature increase with stellar effective temperature. Except for AB\,Aur, the
bulk of the CO gas is at temperatures smaller than 17 K, below the condensation temperature on grains. We
find the scale height of the CO distribution to be larger (by 50\%) than the expected hydrostatic scale
height. The total amount of CO and the isotopologue ratio depends globally on the star. The more UV luminous
objects appears to have more CO, but there is no simple dependency. The [\tco]/[\hco] ratio is $\sim 600$,
with substantial variations between sources, and with radius.
The temperature behaviour is consistent with expectations, but published chemical models have difficulty
reproducing the observed CO quantities. Changes in the slope of the surface density distribution of CO,
compared to the continuum emission, suggest of a more complex surface density distribution than usually
assumed in models. Vertical mixing seems an important chemical agent, as well as photo-dissociation by the
ambient UV radiation at the disk outer edge. }
{}


\maketitle{}

\footnotetext[1]{Based on observations carried out with the IRAM
Plateau de Bure Interferometer. IRAM is supported by INSU/CNRS
(France), MPG (Germany) and IGN (Spain).}
\section{Introduction}

   CO rotation line observations of low-mass and intermediate-mass
Pre-Main-Sequence (PMS) stars in Taurus-Auriga ($\sim140$~pc, Kenyon \etal\ 1994) provide strong evidences
that both TTauri and Herbig Ae stars are surrounded by large ($R_\mathrm{out}\sim 200-800$~AU) Keplerian
disks (GM\,Aur: \citet{Koerner_etal1993}, GG\,Tau: \citet{Dutrey_etal1994}, MWC\,480:
\citet{Mannings_etal1997}). Differences between TTauri and Herbig Ae disks are observed in the very inner
disk, very close to the star \citep{Dullemond_etal2001,Monnier_Millan2002}. However, detailed comparisons,
namely based on resolved observations instead of SED analysis, between Herbig Ae disks and TTauri disks are
too few to allow conclusions about the whole disk. In particular, there is no specific study on the influence
of the spectral type on the disk properties. This can be achieved by comparing the properties of outer disks
imaged by millimeter arrays.

The first rotational lines of CO (J=1-0 and J=2-1) mostly trace the cold gas located in the outer disk ($R>
30$ AU) since current millimeter arrays are sensitivity limited and do not allow the detection of the warmer
gas in the inner disk. In the last years, several detailed studies of CO interferometric maps have provided
the first quantitative constrains on the physical properties of the outer disks. Among them, the method
developed by \citet{Dartois_etal2003} is so far the only way to estimate the overall gas disk structure.
\citet{Dartois_etal2003} deduced the vertical temperature gradient of the outer gas disk of DM\,Tau from a
multi-transition, multi-isotope analysis of CO J=1-0 and J=2-1 maps. The dust distribution and its
temperature can be traced by IR and optical observations and SED modelling
\citep{DAlessio_etal1999,Dullemond_etal2004}. However \citet{Pietu_etal2006} recently demonstrated that SED
analysis are limited by the dust opacity. Due to the high dust opacity in the optical and in the IR, only the
disk surface is properly characterized. Images in the mm domain, where the dust is essentially optically
thin,  revealed in the case of MWC~480 a dust temperature which is significantly lower than those inferred
from SED analysis and from CO images. With ALMA, the method described in this paper, which is today only
relevant to the outer disk, will become applicable to the inner disk of young stars.

In this paper, we focus on the gas properties deduced from CO data. The analysis of the millimeter continuum
data, observed in the meantime, is presented separately \citep{Pietu_etal2006}).  Here, we generalize the
method developed by \citet{Dartois_etal2003} and we apply it to two Keplerian disks surrounding the TTauri
star LkCa\,15 \citep[spec.type K5,][]{Duvert_etal2000} and the Herbig Ae star MWC\,480 \citep[A4 spectral
type,][]{Simon_etal2000}. By improving the method, we also discuss the estimate of disk scale heights. Then,
we compare their properties with those surrounding Herbig Ae and TTauri disks such as AB\,Aur
\citep{Pietu_etal2005}, HD\,34282 \citep{Pietu_etal2003}, DM\,Tau \citep{Dartois_etal2003}.

After presenting the sample of stars and the observations in Sect. 2, we describe the improved analysis
method in Sect. 3. Results are presented in Sect. 4 and their implications discussed in Sect. 5.

\section{Sample of Stars and Observations}

\subsection{Sample of Stars}

The stars were selected to sample the wider spectral type range with only a few objects. Table \ref{tabcoord}
gives the coordinates and spectral type of the observed stars, as well as of HD\,34282 and GM\,Aur which were
observed in \dco\ by \citet{Pietu_etal2003} and \citet{Dutrey_etal1998} respectively. Our stars have spectral
type ranging from M1 to A0. All of them have disks observed at several wavelengths and are located in a hole
or at the edge of their parent CO cloud.

\paragraph{DM\,Tau:}
DM\,Tau is a {\it bona fide} T Tauri star. \citet{Guilloteau_Dutrey1994} first detected CO lines indicating
Keplerian rotation using the IRAM 30 meter telescope. \citet{Guilloteau_Dutrey1998} have confirmed those
findings by resolving the emission of the \dco~\juz~emission using IRAM Plateau de Bure interferometer (PdBI)
and derived physical parameters of its protoplanetary disk and mass for the central star.
\citet{Dartois_etal2003} performed the first high resolution multi-transition, multi-isotope study of CO, and
demonstrated the existence of a vertical temperature gradient in the disk.

\paragraph{LkCa\,15:}
This is a CTT isolated from its parent cloud. \citet{Duvert_etal2000} and \citet{Qi_etal2003} observed the
disk in \dco\ and in HCO$^+$. High resolution \dco\ data was obtained by \citet{Simon_etal2000} who used it
to derive the stellar mass.

\paragraph{MWC\,480:}
The existence of a disk around this A4 Herbig Ae star was first reported by \citet{Mannings_etal1997}. Using
data from OVRO, they perform a modelling of the gas disk consistent with Keplerian rotation. More recently,
\citet{Simon_etal2000} confirmed that the rotation was Keplerian and measured the stellar mass from the CO
pattern.

\paragraph{AB\,Aurigae:}
AB\,Aur is considered as the proto-type of the Herbig Ae star (of spectral type A0) which is surrounded by a
large CO and dust disk. The reflection nebula, observed in the optical, extends far away from the star
\citep{Grady_etal1999}. \citet{Semenov_etal2005} used the 30-m to observe the chemistry of the envelope.
\citet{Pietu_etal2005} has analyzed the disk using high angular resolution images of CO and isotopologues
from the IRAM array and found that, surprisingly, the rotation is not Keplerian. A central hole was detected
in the continuum images and in the more optically thin \tco\ and \cdo\ lines.


\begin{table*}[ht]
\caption{Star properties}\label{tabcoord}
\begin{center}
\begin{tabular}{lllllll}
\hline
Source   & Right Ascension      & Declination & Spect.Type & Effective Temp.(K) & Stellar Lum.($\lsun$) & CO paper \\
\hline \hline
LkCa\,15  & 04:39:17.790  & 22:21:03.34 & K5   &  4350 & 0.74 & 1,2\\
MWC\,480  & 04:58:46.264  & 29:50:36.86 & A4   &  8460 & 11.5 & 1,2\\
DM\,Tau   & 04:33:48.733  & 18:10:09.89 & M1   &  3720 & 0.25 & 3  \\
AB\,Aur   & 04:55:45.843  & 30:33:04.21  & A0/A1& 10000 & 52.5 & 5  \\
HD 34282 & 05:16:00.491  &-09:48:35.45 & A1/A0&  9440 & 29   & 4  \\
GM Aur   & 04:55:10.98    & 30:21:59.5   & K7   &  4060 & 0.74 & 6  \\
 \hline
\end{tabular}
\end{center}
{Col.~2,3: J2000 coordinates deduced from the fit of the 1.3\,mm continuum map of the PdBI. Errors bars on
the astrometry are of order $\leq 0.05''$. Col.~4,5, and 6, the spectral type, effective temperature and the
stellar luminosity are those given in \citet{Simon_etal2000} and \citet{Pietu_etal2003} for HD\,34282 and
\citet{VandenAncker_etal1998} for AB\,Aur. Col.6, list of CO interferometric papers are: 1 = this paper, 2 =
\citet{Simon_etal2000}, 3 = \citet{Dartois_etal2003}, 4 = \citet{Pietu_etal2003}, 5 = \citet{Pietu_etal2005},
6 = \citet{Dutrey_etal1998}.}
\end{table*}

\subsection{PdBI data}

For DM\,Tau, LkCa\,15, and MWC\,480, the $^{12}$CO J=2-1 data were observed in snapshot mode during winter
1997/1998 in D, C2 and B1 configurations. Details about the data quality and reduction are given in
\citet{Simon_etal2000} and in \citet{Dutrey_etal1998}.  The $^{12}$CO J=2-1 data were smoothed to 0.2 $\kms$
spectral resolution. The HCO$^+$~\juz\ data were obtained simultaneously: a 20 MHz/512 channels correlator
unit provided a spectral resolution of 0.13 $\kms$. The $^{13}$CO J=1-0 and J=2-1 observations of DM\,Tau are
described in \citet{Dartois_etal2003}.

We obtained complementary $^{13}$CO J=2-1 and J=1-0 (together with C$^{18}$O J=1-0 observations) of MWC\,480
and LkCa\,15. These observations were carried out in snapshot mode (partly sharing time with AB\,Aur) in
winter periods between 2002 and 2004. Configurations D, C2 and B1 were used at 5 antennas. MWC\,480 has also
shortly been observed in A configuration simultaneously with AB\,Aur. Finally, data was obtained in the
largest A$^+$configuration for LkCa\,15 and MWC\,480 in January 2006 \citep[for a description of these
observations, see][]{Pietu_etal2006}. Both at 1.3\,mm and 2.7\,mm, the tuning was double-side-band (DSB). The
backend was a correlator with one band of 20 MHz, 512 channels (spectral resolution 0.11 $\kms$) centered on
the $^{13}$CO J=1-0, one on the C$^{18}$O J=1-0 line, and a third one centered on the \tco~\jdu\ line
(spectral resolution 0.05 $\kms$), and 2 bands of 160 MHz for each of the 1.3 mm and 2.7 mm continuum. The
phase and secondary flux calibrators were 0415+379 and 0528+134. The rms phase noise was 8$\degr$ to
25$\degr$ and 15$\degr$ to 50$\degr$ at 3.4 mm and 1.3 mm, respectively (up to 70$\degr$ on the 700 m
baselines), which introduced position errors of $\leq 0.05\arcsec$. The estimated seeing is about 0.2$''$.

Flux density calibration is a critical step in such projects. The kinetic temperature which is deduced from
the modelling is directly proportional to the measured flux density. This is even more problematic since we
compare data obtained on several years. Hence, we took great care of the relative flux calibration by using
not only MWC\,349 as a primary flux reference, assuming a flux S$(\nu) = 0.95 (\nu/87 \mathrm{GHz})^{0.60}$
(see the IRAM flux report 14) but also by having always our own internal reference in the observations. For
this purpose we applied two methods: i) we used MWC\,480 as a reference because its continuum emission is
reasonably compact and bright, ii) we also checked the coherency of the flux calibration on the spectral
index of DM\,Tau, as described in \citet{Dartois_etal2003}. We cross-checked all methods and obtained by
these ways a reliable relative flux calibration from one frequency to another.

\figMAPS
\subsection{Data Reduction and Results}

We used the GILDAS\footnote{See \texttt{http://www.iram.fr/IRAMFR/GILDAS} for more information about the
GILDAS softwares.}  software package \citep{pety05} to reduce the data and as a framework to implement our
minimization technique. Figure \ref{fig:maps} presents some channel maps of \dco, \tco~\jdu\ and \tco~\juz\
for LkCa\,15 and MWC\,480. Similar figures for DM\,Tau can be found in \citet{Dartois_etal2003}, and for
AB\,Aur in \citet{Pietu_etal2005}. A velocity smoothing to 0.6 $\kms$ (for MWC\,480) and 0.4 $\kms$ (for
LkCa\,15) was applied to produce the images displayed in Fig.\ref{fig:maps}. However, in the analysis done by
minimization, we used spectral resolutions of $0.20 \kms$ (\dco\ and HCO$^+$) and 0.21 $\kms$ (\tco), and
natural weighting was retained.

\figHCO

The HCO$^+$ images are presented in Fig.\ref{fig:hco}. The typical angular resolution of these images is $3 -
4''$. For LkCa\,15, the line flux is compatible with the result quoted by \citet{Duvert_etal2000} and
\citet{Qi_etal2003}, and for DM\,Tau with the single-dish measurement of \citet{Dutrey_etal1997}.

\section{Model description and Method of Analysis}
\label{sec:model}

In this section, we summarize the model properties and describe the
improvements since the first version \citep{Dutrey_etal1994}.

\subsection{Solving the Radiative Transfer Equation}

To solve the radiative transfer equation coupled to the statistical equilibrium, we have written a new code,
offering non LTE capabilities. More details are given in \citet{Pavlyuchenkov2007}.
The LTE part, which is sufficient for low energy level of rotational CO lines (J=1-0 and J=2-1), is strictly
similar to that of \citet{Dutrey_etal1994}, and uses ray-tracing to compute the images. The HCO$^+$ data was
also analyzed in LTE  mode (see Sect.\ref{sec:temp} for the interpretation of this assumption). We use a
scheme of nested grids with regular sampling to provide sufficient resolution in the inner parts of the disk
while avoiding excessive computing time.

\subsection{Description of the physical parameters}

For each spectral line, the relevant physical quantities which control the line emission are assumed to vary
as power law as function of \textit{radius} (see Sect.\ref{sec:cylindre} for the interpretation of this term)
and, except for the density, do not depend on height above the disk plane. The exponent is taken as positive
if the quantity decreases with radius:
 $$ a(r) = a_0 (r/R_a)^{-e_a}$$ For each molecular line, the disk is
thus described by the following parameters :
\begin{itemize}\itemsep 0pt
 \item $X_0, Y_0$, the star position, and $V_\mathrm{disk}$, the
 systemic velocity.
 \item PA, the position angle of the disk axis, and $i$ the
 inclination.
 \item $V_0$, the rotation velocity at a reference radius $R_v$, and
 $v$ the exponent of the velocity law. With our convention, $v = 0.5$
 corresponds to Keplerian rotation. Furthermore, the disk is oriented
 so that the $V_0$ is always positive. Accordingly, PA varies between
 0 and $360^\circ$, while $i$ is constrained between $-90^\circ$ and
 $90^\circ$ (see fig. \ref{fig:geo}).
 \item $T_\mathrm{m}$ and $q_m$, the temperature value at a reference
 radius $R_T$ and its exponent (see Sect.\ref{sec:temp}).
 \item $dV$, the local line width, and its exponent $e_{v}$. The
 interpretation of $dV$ is discussed in Sect.\ref{sec:dv}
 \item $\Sigma_m$, the molecular surface density at a radius $R_\Sigma$
 and its exponent $p_m$
 \item $R_\mathrm{out}$, the outer radius of the emission, and
 $R_\mathrm{in}$, the inner radius.
 \item $h_m$, the scale height of the molecular distribution at a
 radius $R_h$, and its exponent $e_h$: it is assumed that the density
 distribution is Gaussian, with
 \begin{equation}\label{eq:scale} n(r,z) = \frac{\Sigma(r)}{h(r) \sqrt{\pi}}
 \exp\left[-\left(z/h(r)\right)^2\right] \end{equation}
 (note that with this definition, $e_h < 0$ in realistic disks)
 \end{itemize}
thus giving a grand total of 17 parameters to describe the
emission.

It is important to realize that all these parameters can actually be constrained for the lines we have
observed, under the above assumption of power laws. This comes from two specific properties of
proto-planetary disks: i) the rapid decrease of the surface density with radius, and ii) the known kinematic
pattern. The only exception is the inner radius, $R_\mathrm{in}$, for which an upper limit of typically 20 --
30 AU can only be obtained.

The use of power laws is appropriate for the velocity, provided the disk self-gravity remains small. Power
laws have been shown to be a good approximation for the kinetic temperature distribution \citep[see,
e.g.][]{Chiang_etal1997}, and thus to the scale height prescription. For the surface density, power laws are
often justified for the mass distribution based on the $\alpha$ prescription of the viscosity with a constant
accretion rate: self-similar solutions to the disk evolution imply power laws with an exponential edge
\citep{Hartmann_etal1998}. Although \citet{Malbet_etal2001} pointed out some limitations of this assumption
for the inner parts ($r < 30$ AU) of the disk, the approximation remains reasonable beyond 50 AU. However,
although this may be valid for H$_2$, chemical effects may lead to significant differences for any molecule.
Given the limited spatial dynamic range and signal to noise provided by current (sub-)millimeter arrays,
using more sophisticated prescriptions is not yet justified, and power laws offer a first order approximation
of the overall molecular distribution.

\convention

\tabCylSphere

\subsubsection{Spherical vs Cylindrical representation}
\label{sec:cylindre}

To describe the disk and its parameters, we can use either cylindrical
or spherical coordinates.

The description given in the previous sub-section is unambiguous for thin disks (where $h(r) \ll r$),
However, for molecular disks with outer radius as large as 800 AU, the disk thickness is no longer small, and
the classical theory of hydrostatic scale height, which uses cylindrical coordinates, is only valid in the
small $h/r$ limit. It may then become significant to distinguish between a \textit{spherical} representation,
where the functions depends on the distance to the star ($r=d$) and height above disk plane, $z$
  \begin{equation}\label{eq:sphere} F = f(d,z) \end{equation}
and a \textit{cylindrical} representation, where the functions depends on the projected distance on the disk
($r=\rho$) and height $z$
  \begin{equation}\label{eq:cylindre} F = f(\rho,z) \end{equation}
with $d^2 = \rho^2+z^2$.  It is obvious that the exponent of the power laws will depend slightly on which
description is selected. We have evaluated the differences between the two different representations (see
Tab.\ref{tabcyl-spher}): as expected, the results are very similar for ~\tco, but the cylindrical
representation is a $\sim$ 8 sigma better representation for ~\dco~\jdu~ (the most affected transition
because of its higher opacity). The most affected parameters are $\Sigma$ and $h$. However, with the
exception of the scale height, the effects remain small compared to the uncertainties. Accordingly, we have
arbitrarily decided to represent all parameters in terms of \textit{cylindrical} laws, i.e. the radius is $r
= \rho$ in the model description.

\subsubsection{Reference Radius}

Unless $p_m \simeq 0$, the choice of the reference radius $R_\Sigma$ will affect the relative error bar on
$\Sigma_m$, $\delta \Sigma_m / \Sigma_m$. Depending on the angular resolution and on the overall extent of
the emission, for each molecular line, there is a different optimal radius $R_\Sigma$ which minimizes this
relative error. The same is true for all other reference radii for the power laws. This makes direct
comparison at an arbitrary radius not straightforward. To circumvent this problem, we have determined
$\Sigma_m$, $\delta \Sigma_m$ as a function of the reference radius for each transition, and selected the
$R_{\Sigma}$ giving the best S/N ratio. Results are given Fig.\ref{fig:raddm}-\ref{fig:radmw}: note that the
curves $\Sigma_m(R_\Sigma)$ should not be interpreted as independent estimates of the local surface density
at different radii: they explicitly rely on the assumption of a single power law throughout the whole disk.

The same arguments can be applied to all other power laws, in particular to the scale height and the
temperature. However, for the latter, the exponent is usually small ($q = -0.2 - 0.5$), so the choice of the
reference radius is less critical. Figure \ref{fig:raddm} shows the variation of the parameters (here
$\Sigma_m$, $T_m$ and $h$) and of the signal to noise ratio as a function of the reference radius for
DM\,Tau. This is a reanalysis of the data published by \citet{Dartois_etal2003}. This method gives consistent
results with previous work, but allows a more precise determination of the parameters (especially the surface
density as stated above), by determining in which region they are constrained. Fig.\ref{fig:raddm} indicates
that the temperature is determined around 100 -- 200 AU, while the surface densities are constrained in the
300 -- 500 AU region.  Somewhat smaller values apply to the other sources, LkCa\,15 and MWC\,480 (see
Fig.\ref{fig:radlk}-\ref{fig:radmw}).

\subsubsection{Temperature law}
\label{sec:temp} The interpretation of $T_m$ (and as a consequence of $\Sigma_m$) depends on the specific
model of radiative transfer being used. In this paper, we are dealing with CO isotopologues, for which LTE is
good approximation: $T_m$ is then the kinetic temperature. For transitions which may not be thermalized, two
different radiative transfer models may be used. We can apply an LTE approximation: since we are fitting
brightness distributions, $T_m$ is in this case the excitation temperature $T_\mathrm{ex}$, and the surface
density $\Sigma_m$ is computed assuming the same temperature controls all level population, i.e. the
partition function.  We can also solve for molecular line excitation using a non-LTE statistical equilibrium
code: $T_m$ is then the kinetic temperature, and $\Sigma_m$ the total molecule surface density, within the
limitations of the radiative transfer code accuracy. In this paper, only the LTE mode was used.

\subsubsection{Surface Density}
The surface density is expected to fall off rapidly with distance from the star ($p(\mathrm{H}_2) \simeq 1 -
2$), and in particular much more rapidly than the temperature ($q\simeq 0.5$). Remembering that, because of
the partition function, the line opacity of a $J=1-0$ transition scales as $\Sigma_m/T^2$ at high enough
temperatures (for constant line width), this indicates that, in general, the central part of the disk is much
more optically thick than the outer regions. For the detectable lines, the temperature can be derived from
the emission from this optically thick core, while the surface density $\Sigma_m$ is derived from the
optically thin region. This remains valid provided the same temperature law applies to the two regions. For
the weakest lines, the optically thick core may be too small, and the temperature exponent remains
essentially unconstrained: this should be reflected to by the error bars on the temperature. In such cases,
however, the line emission scales as $\Sigma_m/T$ (for J=1-0 line).

If the molecule is very abundant, such as CO, the local line core may be optically thick almost throughout
the disk. However, the local line wings remain optically thin: if the velocity resolution is high enough,
this information can be used to constrain the opacity, and hence the surface density, although less
accurately. To illustrate this point, let us consider the case of a face on disk. The apparent width of the
line profile is then controlled only by the internal velocity dispersion (turbulent + thermal) and by the
line opacity. In the outer parts, the line is optically thin and of Gaussian shape, while in the inner parts,
the line is broadened by the large opacity, and the profile tends towards a square shape, with width
proportional to $\sqrt{\ln(\tau)}$. The evolution of the line profiles as function of radius indicates at
which radius $\tau=1$, and thus constrains the molecular surface density. Note that if the spectral
resolution is insufficient to properly sample the line width, this is no longer possible. Current correlators
are not instrumentally limited but smoothing can be necessary when the sensitivity is the limiting factor. If
the disk is not seen face on, the systematic velocity gradient get superimposed to this local line
broadening, but does not change the fundamental relationship between the apparent (local) line width and the
opacity.  In any case, since the effect goes as $\sqrt{\ln(\tau)}$, measuring the surface density by such a
method is expected to be rather unprecise, and will result in larger errors. Furthermore, it should be noted
that this opacity broadening effect results in a slight coupling of the line width parameters, $dV$ and
$e_v$, to the surface density.

\subsubsection{Scale Height}
\label{sec:scale}

The existence of a known kinematic pattern also allows to constrain the scale height parameters $h_0$ and
$e_h$. Unless the disk is seen face on, for an optically thin line, the velocity range intersected along a
line of sight will depend on the disk inclination and flaring, resulting in a coupling between scale height
and line width. For an optically thick line, the difference in inclinations between the front and back
regions of the disk can also be measured. The spatial distribution of the emission as a function of projected
velocity (or function of the velocity channels) along the line of sight will depend on the scale height of
the disk (this is illustrated for the opacity by Fig.3 from \citet{Dartois_etal2003}). At first glance, the
expected scale heights, of order 50 AU at 300 AU from the star, would appear too small compared to the
resolution of $0.7 - 1.5''$ (100 to 200 AU), and the exponent $e_h$ even more difficult to constrain.
However, millimeter interferometry, being an heterodyne technique, allows phase referencing between velocity
channels, and the precision in relative positions is equal to the spatial resolution divided by the Signal to
Noise (as our bandpass accuracy, about a degree, is not the limiting factor), i.e. $\sim 10$~AU. This
super-resolution approximately matches the expected displacements due to Keplerian rotation,
\begin{equation}\label{eq:super}
 \delta r = 2 r \frac{\delta v}{v} \simeq 10~\mathrm{AU~at}~r = 100~\mathrm{AU}
\end{equation}
for $\delta v = 0.15 \kms$. Note that in Eq.\ref{eq:super}, the value to be used for $\delta v$ is
approximately the largest of the local line width and the spectral resolution. As a result, the scale height
has a measurable effect on the images, even though the flaring parameter is difficult to constrain.

Actually, because of the effect of the scale height on the disk images, if a simple hydrostatic equilibrium
is assumed instead of handling the scale height separately, it will result in a bias in the temperature law
since one is then introducing a coupling between the vertical distribution of the molecules and the
temperature which is vertically isothermal in our model.

Finally, note that because the derivation of the scale height depends on the differing inclinations of the
optically thick parts, there is also a coupling between the surface density and the scale height derivation.
This coupling is specially strong for disks close to edge-on.

\subsubsection{Line Width}
\label{sec:dv}

Line broadening results from a combination of thermal and turbulent velocities.
\begin{equation}
\Delta V = \sqrt{\frac{2kT}{m} + v_\mathrm{turb}^2}
\label{eq:turb1}
\end{equation}
However, an accurate representation of the thermal broadening requires a correct fitting of the gas
temperature. This can become problematic when dealing with lines which are sub-thermally excited.
Accordingly, our code offers two different parameterizations of the line width. The first one is
\begin{equation}
\Delta V = \sqrt{\frac{2kT}{m} + dV^2} \mathrm{~~i.e.~~} dV =
v_\mathrm{turb} \label{eq:turb2}
\end{equation}
which corresponds, for $e_v = 0$ ($v_\mathrm{turb}= cte$), to the
parametrization used in our previous papers.  The second one is
\begin{equation}
\Delta V = dV = dV_m (r/R_{dV})^{- e_v} \label{eq:turb3}
\end{equation}
i.e., we do not separate the thermal component from the total line width. This offers the advantage of
providing values which are not biased when the kinetic temperature is difficult to constrain (as can happen
for lines which are not at LTE). We use this new description in this paper.

Note that all line widths we quote are half-width at $1/e$. This allows direct comparison with the thermal
velocity, $\sqrt{2kT/m}$, but one should multiply by $2\sqrt{\ln(2)} = 1.66$ to convert to FWHM. As described
before, the line width is also coupled to the scale height.

\radiusDMTAU

\radiusLKCA

\radiusMWC

\subsection{Minimization Technique}
Using the disk parametrization described above, we have improved several aspects of the method originally
developed by \citet{Dartois_etal2003}.

\subsubsection{Minimization Method and Error Bars}

The comparison between model and data and the $\chi^2$ minimization is always performed inside the UV plane,
using natural weighting \citep{Guilloteau_Dutrey1998}.  We have implemented a modified Levenberg-Marquardt
minimization scheme to search for the minimum, and we use the Hessian to derive the error bars.  This scheme
is much faster than the grid search technique used previously, and less prone to local minima. It allows to
fit simultaneously more parameters, and thus provide a better estimates of the errors because the coupling
between parameters is taken into account.

However, the current fit method does not handle asymmetric error bars which happen in skewed distributions:
this limitation should not be ignored when considering the errors on $\Sigma$, $T$ or $R_\mathrm{out}$. The
quoted error bars include thermal noise only, but do not include calibration errors (in phase and in
amplitude). Amplitude calibration errors will directly affect the absolute values of the temperature or the
surface density (as a scaling factor, but not the exponent) while phase errors can introduce a seeing effect.
The latter effect is very small for line analysis because molecular disks are large enough but it starts to
be significant for dust disks. Our careful calibration procedure brings the amplitude effect to below 10 \%.
All other parameters are essentially unaffected by calibration errors.

\subsubsection{Continuum Handling}
\label{sec:fitcont} The speed improvement allowed us to treat more properly the continuum emission. Even
though the continuum brightness is small compared to the line brightness, continuum emission cannot be
ignored when trying to retrieve parameters from the line emission because it appears systematically in all
spectral channels.

A simultaneous fit of the continuum emission with its own physical parameters will properly take into account
these biases. We performed this by adding to the spectral $UV$ table a channel dedicated to the continuum
emission. The global $\chi^2$ is given by the sum $\chi^2 = \chi^2_C + \chi^2_L$ where $\chi^2_C$ and
$\chi^2_L$ are the $\chi^2$ for the continuum and for the line minimization, respectively. The continuum
emission is fitted using the following parameters: $R_{Cint}$, $R_{Cout}$, the inner and outer radii of the
dust distribution, $\Sigma_d$ and $p_d$, the surface density power law, and $T_d$ and $q_d$, the dust
temperature power law. Note that the representation of the continuum needs only be adequate within the noise
provided by the spectral (total) line width, about 10 to 20 MHz, rather than for the whole receiver bandwidth
(600 MHz). A simplified model is thus often acceptable, as shown below.

We also compared this robust method with a simplified (and faster) one in which we subtract, inside the $UV$
plane, the continuum emission from the line emission before fitting. This subtraction is not perfect: for the
optically thick regions (which always occupy some fraction of the line emission) continuum should not be
subtracted. However, unless the disk is seen face on, because of the Keplerian shear, at each velocity, the
line emitting/absorbing region only occupies a small fraction of the disk area. This fraction is at most of
order of the ratio of the local line width to the projected rotation velocity at the disk edge, in practice
less than 10 -- 15 \% for the disks we considered. Accordingly, the error made by subtracting the whole
continuum emission in every channel is quite small.

We verified that this simplified method gives, as expected, the same results as the robust one. We used it,
as it is significantly faster. This better handling of the continuum allowed us to improve on the mass
determination of MWC\,480 \citep{Simon_etal2000}, where the continuum is relatively strong.

\subsubsection{Multi-Line fitting}

The code has also been adapted to allow minimization of more than one data set at a time. This has been used
to simultaneously fit the \tco~\juz\ and \jdu\ transitions. For multi-line fitting, the temperature is
derived from the line ratio. The temperature is determined in the disk region where both lines are optically
thin. In this case, the fit is much more accurate than from a single line observation if there is no
significant vertical temperature gradient.  When a temperature gradient is present, the derived temperature
will reflect an ``average'' temperature weighted by the thermal noise.

\section{Results}

\subsection{Interpretation of the Power Law model}

The parametric model would perfectly describe the line emission from a molecule in LTE in a vertically
isothermal disk which is in hydrostatic equilibrium, with power law for the kinetic temperature and surface
density, and constant molecular abundance. In such a case, $e_v = q/2$ if the local line width is thermal,
and $h = q/2 - 1 - v$ and $h_0 = (\sqrt{2 k T_0/\overline{m}}) / V_0$ (using the same reference radius $R_h =
R_T = R_v$, $\overline{m}$ being the mean molecular weight). If such a disk was chemically homogeneous, all
molecular lines would yield the same results for the 17 parameters (after correction of $\Sigma_m$ for the
molecule abundance). Differences between the parameters derived from several transitions will reflect
departures from such an ideal situation.

\geom

For example, the geometric parameters PA and $i$ should all be identical, as should be the kinematic
parameters $V_\mathrm{disk}, V_0, v$ (and we should have $v =0.50$ for Keplerian rotation). $X_0, Y_0$ should
reflect the absolute astrometric accuracy of the IRAM interferometer. Different values for $T_m$ from several
transitions for the same molecule, or its isotopologues, may reveal vertical temperature gradients, or, in
case of non-LTE excitation, density gradients, since such transitions probe different regions of the disk.
The values of $T_m$ from different molecules may also provide constraints on the density, because of the
different critical densities. And, more directly, the values of $\Sigma_m$, $p_m$, $R_\mathrm{in}$ and
$R_\mathrm{out}$ will reflect the chemical composition of the disk as function of radius.

\subsection{Geometric and kinematic parameters}

Figure \ref{fig:para} shows the geometric and kinematic parameters PA, $i$ and rotation velocity $V_{100}$
derived from the observed transitions in DM\,Tau, LkCa\,15 and MWC\,480. The grey area represent the 1
$\sigma$ range of the mean value computed from the CO lines. As expected, the derived value are in very good
agreement altogether. In particular, the distribution of the derived values and the error bars are consistent
with a statistical scattering, meaning that: i) the observed lines come from the same disk, ii) the error
bars have the good magnitude, and iii) there is no bias in the determination of the parameters.
This analysis of the ``basic'' parameters is of prime interest, because it shows that the analysis method is
robust, and that the error bars have a real physical meaning. Although noisier, the kinematic and geometric
parameters determined from \hco~ are in agreement with those found from CO.  We thus used the CO-derived
values to determine the surface density and excitation temperature of \hco{}.

\subsection{Results}

Tables \ref{tab:dmtau}, \ref{tab:lkca} and \ref{tab:mwc480}, present the best fit results for DM\,Tau,
LkCa\,15 and MWC\,480, respectively. Compared to previous papers
\citep[e.g.][]{Simon_etal2000,Duvert_etal2000,Dutrey_etal1998}, the new fits include all the refinements of
the method described in Sec.\ref{sec:model} and in \citet{Pietu_etal2006}.

All quoted results where obtained using cylindrical coordinates and the continuum was subtracted to the
spectroscopic data before analysis. All the models were obtained in the same manner. In particular, we
assumed a constant line width ($e_v = 0$), Keplerian rotation ($v = 0.5$), and a flaring exponent $e_h =
-1.25$ (the value for a Keplerian disk in hydrostatic equilibrium with a temperature radial dependence of
0.5). This value is very close to those found by \citet{Chiang_etal1997} for the super heated layer model.
Systematic tests with different sets of free parameters ensure that the geometrical and kinematical
parameters are insensitive to these assumptions. The temperature $T_m(r)$ is also hardly affected.  The
surface density $\Sigma_m(r)$ and outer radius $R_{out}$ depends to some extent on the assumed flaring
parameter $e_h$. However, because of the coupling between scale height and line width, both $H_0$ and $dV$
should be taken with some caution. The dynamical mass derived from \dco\ also depends weakly on the assumed
scale height, since we derive the inclinations from the apparent CO surface.

The differences between the values reported here for DM\,Tau and those published by \citet{Dartois_etal2003}
come from the different assumptions in the analysis. \citet{Dartois_etal2003} assumed hydrostatic
equilibrium, and model the line width following Eq.\ref{eq:turb2} . They also assumed [\dco]/[\tco] = 60.

\TableDM

\TableLKCA

\TableMWC

The \tco\ emission in MWC\,480  is strong enough to allow separate fitting of the \jdu\ and \juz\ lines. We
thus present separate results for each line. We also present a result in which both lines are fitted
together: as expected in this case, the derived temperature is in between the results from the \jdu\ and
\juz\ transitions. However, for LkCa\,15, the \tco\ emission is too weak for a separate analysis: only the
results from the simultaneous fitting of both \tco\ lines is presented.

For \hco\, in MWC\,480, the signal to noise did not allow an independent derivation of the temperature $T_m$:
we used the law derived from \tco . A similar problem occurs in LkCa\,15: for simplicity, we used $T_m =
19$~K and $q_m = 0.38$, values compatible with the CO results, although there is a solution marginally better
($1.7 \sigma$) with slightly lower temperatures ($T_m = 13$~K), with a somewhat steeper surface density
gradient.

\section{Discussion}

We discuss here the physical implications of the results presented in the preceding section. The following
subsections focus on the measurement of star masses, of the vertical temperature gradient, of the hydrostatic
scale heights, of the CO abundances and CO outer radii, and of the HCO$^+$ distribution.

\subsection{CO dynamical masses}

Table \ref{tab:mass} shows a comparison of the mass determination obtained in this work and in
\citet{Simon_etal2000}. The subtraction of the continuum has slightly improved the mass determination in the
case of MWC\,480. \citet{Simon_etal2000} used the old Solar metallicity ($Z=0.02$) pre main-sequence (PMS)
models. In Fig.\ref{fig:hrd-mass}, we present here also results for a metallicity $Z=0.01$, much closer to
the new Solar metallicity \citep[$0.0126$, see][and references therein]{Asplund_etal2004,Grevesse_etal2005}.
LkCa\,15 is in agreement with models of both metallicities ($M=0.95\pm0.05\msun,t=5-6$ Myr and
$1.1\pm0.1\msun, t=4-6$Myr, for $Z=0.01$ and $Z=0.02$ respectively). In fact, the measured dynamical mass
does not provide any strong constraint on the location of LkCa\,15 in this diagram (since the PMS tracks are
almost parallel to the Y axis for stars with $0.5<M_*/M_\odot<1$  in this distance independent HR diagram),
but serves as a consistency check. However for MWC\,480, either the metallicity is $Z=0.01$ (i.e. nearly
solar), in which case the derived dynamical mass is in agreement with the \citet{Siess_etal2000} tracks
($M=1.8 \msun,t=8$ Myr), or the metallicity is 0.02 and MWC\,480 would be located at a larger distance as
suggested by \citet{Simon_etal2000} (mass $1.9-2.0\msun$, age $\sim7$ Myr, and the distance would need to be
increased by about 10 \%)

\begin{table}[!ht]
 \caption{Stellar masses}\label{tab:mass}
\begin{tabular}{lcc}
 \hline
 \hline
 Source & Previous mass & This work \\
 & ($\msun$) & ($\msun$) \\
 \hline
 DM\,Tau  & $0.55\pm 0.03$ & $0.53 \pm 0.02$\\
 LkCa\,15  & $0.97\pm 0.03$ & $1.01 \pm 0.03$\\
 MWC\,480 & $1.65\pm 0.07$ & $1.83 \pm 0.05$\\
\hline
 \end{tabular}\\
 {See Simon \etal 2000 for the previous measurements. The old masses
 are based only on \dco\ while the new measurements are the average of
 all available CO lines.}
 \end{table}

\FigMASS

\subsection{Vertical Temperature Gradient}

The primary goal of these observations was to confirm the findings of \citet{Dartois_etal2003}: the existence
of a vertical temperature gradient. Fig.\ref{fig:temp} shows the temperatures deduced from the various CO
lines versus the star effective temperatures in all sources of the sample. We confirm the DM\,Tau results
with this slightly different analysis. MWC\,480 also presents a significant temperature difference between
\dco\ and \tco. However, in LkCa\,15, there is no clear evidence of temperature gradient. As a general trend,
the hotter stars of the sample, namely the Herbig Ae stars, exhibit larger kinetic temperature in the disks,
not only at their surfaces but also for their interiors (see also Sect.\ref{sec:compare} for AB\,Aur). For
MWC\,480, even if the surface appears hotter, a significant fraction of the disk (for $r > 200$~AU) remains
at temperature below the CO freeze-out temperature.

The comparison between the three sources can be understood by noting that the surface densities of \tco\
decreases from DM\,Tau to MWC\,480 and LkCa\,15 (see Sect.\ref{sec:compare}). Moreover, the temperature is
lower in DM\,Tau, resulting in higher opacities for the \jdu\ and \juz\ transitions than in the other
sources. As a result, the location of the $\tau =1$ surface in the \tco\ lines differ in the three sources.
In DM\,Tau, this surface is above (for the \jdu\ line) or at (for the \juz\ line) the disk plane (as seen
from the observer). A similar behavior is observed in MWC\,480.

In LkCa\,15, the \tco\ lines are nearly optically thin throughout the disk, and the temperature is accurately
determined by the \jdu / \juz\ line ratio. The lack of difference between the \dco\ and \tco\ results
indicate both lines are formed in a similar region: there must be little \tco\ at temperatures much lower
than indicated by \dco.

\subsection{CO Surface density and outer radius}
\label{sec:co}

There are significant differences in surface densities between the three sources. At 300 AU, the measured
surface densities of \tco\ range from $\sim 2.5\,10^{15}$ cm$^{-2}$ in LkCa\,15 to $\sim 3.2\,10^{15}$
cm$^{-2}$ in MWC\,480 and $\sim 7\,10^{15}$ cm$^{-2}$ in DM\,Tau. The variations are even larger for \dco\,
with values $\sim 3\,10^{16}$ cm$^{-2}$ in LkCa\,15 and MWC\,480, and 4 times larger, $14\,10^{16}$ cm$^{-2}$
for DM\,Tau (although the latter is uncertain by a factor 2 because of the steep density gradient).

The \dco / \tco\ ratio is $20 \pm 3$ in DM~Tau near 450 AU, the distance where this ratio is best determined.
It is $11 \pm 2$ at 300 AU for LkCa\,15, and $8 \pm 2$ for MWC\,480. These values are much lower than the
standard $^{12}$C/$^{13}$C ratio in the solar neighbourhood. These low values are most likely the result of
significant fractionation, as in the Taurus molecular clouds, since the typical temperatures near 400~AU are
around 12 K (see also Sect.\ref{sec:compare}).

The exponent of the surface density distribution varies significantly. For DM\,Tau and MWC\,480, an exponent
of $\sim 3.5$ is appropriate for all isotopologues. For LkCa\,15, there is a large difference between the
exponent for \tco, 1.5, and that for \dco, 4.5. Each line however samples a very different region of the
disk. The \tco\ data are most sensitive to the 200 -- 300 AU region, while the \dco\ data is influenced by
the outermost parts, 500 -- 700 AU (because of the opacity of the \dco~line) and up to the outer disk radius.
The difference in exponent may thus indicate a steepening of the \dco~surface density distribution at radii
larger than $> 500$~AU. At such radii, the \dco~emission becomes optically thin, and better constrains the
surface density than \tco . For DM\,Tau and MWC\,480, the surface density of CO falls faster with radius than
usually assumed for molecular disks ($p\sim 1-1.5$, for H$_2$). Note that the uncertainty on temperature law
affects only weakly the surface density derivation, specially for the \jdu\ transition for which the emission
is essentially proportional to $\Sigma_m$ for the considered temperature range in the optically thin regime.
This apparent steepening could reflect a change in the (H$_2$) surface density distribution in the disk.
However, it can also be a result of chemical effects. The isotropic ambient UV field will tend to
photo-dissociate CO isotopologues near the disk outer edge. Another effect is that, because of the lower
temperatures, depletion onto dust grains may be more effective at larger radii (although the lower densities
result in a larger timescale for sticking onto grains).

\TableROUT

In all three sources, the outer radius in \tco\ is significantly smaller than in \dco. In LkCa\,15, the large
difference may be partly attributed to the change in the exponent of the density law. However, if we assume
$p = 1.5$ for \dco, we get an outer radius of $710 \pm 10$, still significantly larger than that for \tco .
The effect is thus genuine.

\FigTEMP

\subsection{The LkCa\,15 inner hole}
\label{sec:lkca}

In the case of LkCa\,15, it is also important to mention its peculiar geometry: \citet{Pietu_etal2006} have
discovered a 50 AU cavity in the continuum emission from LkCa\,15. We checked whether there is any hint of
this cavity in the CO data. Adding the inner radius as an additional free parameter yields $R_\mathrm{int} =
13 \pm 5$~AU for \dco\ and $R_\mathrm{int} = 23 \pm 8$~AU for \tco . These values are consistent with a
smaller hole, or no hole at all, in CO and \tco. The 50 AU inner radius determined from the continuum is
excluded at the 7 $\sigma$ level in \dco, and at the 3 $\sigma$ level in \tco.  Moreover the fit of a hole
does not significantly change the value of $p$, for both \dco~and \tco~ emissions.

These small values indicate that the CO gas extends well into the continuum cavity. The cavity is thus not
completely void, but still has a sufficient surface density for the \tco\ lines to be detected. The LkCa\,15
situation is similar to that of AB\,Aur, for which the apparent inner radius increases from \dco\ to \tco\
and to the dust emission \citep{Pietu_etal2005}.

\subsection{\hco\ surface density}

The \hco\ data analysis reveal 4 major facts
\begin{itemize}
 \item the outer radius in \hco\ is similar to that in \dco, and in
 particular it is larger than that in \tco.
 \item in the two cases where it could be determined, the excitation
 temperature of the \hco\juz\ is lower than that of the \tco~\jdu\
 line.
 \item at 300 AU, the [\tco]/[\hco] ratio is $500 \pm 150$ (DM\,Tau),
 $320 \pm 50$ (LkCa\,15), and $1300 \pm 200$ for MWC\,480, with an additional uncertainty around
 50 \% for these last two sources due to the adopted temperature distribution. These values
 are similar to, although slightly smaller than the one found by \citet{Guilloteau_etal1999} for the
 circumbinary disk of GG Tau ($\sim 1500$). On the other end,
 \citet{Pety_etal2006} find a ratio of about 10$^4$ for the low mass
 disk around HH\,30, but at a smaller radius (100 AU).
 \item This ratio varies with radius: there is a tendency to have a
 somewhat flatter distribution of \hco\ than of CO (with the exception
 of \tco\ in LkCa\,15).
 \end{itemize}
In DM\,Tau, the excitation temperature falls down to about 6 K beyond 600 AU, while the CO data indicates
kinetic temperature of about 10 K in this region. Our calibration technique ensures that this is not an
artifact. It may be a hint of very low temperatures in the disk mid-plane. On the other end, it may be an
indication of sub-thermal excitation. However, as the critical density for the \hco~\juz~ is only a few
$10^4$ cm$^{-3}$, such a sub-thermal excitation would require that the \hco\ gas is located well above the
disk plane. Spatially resolved observations of the \hco~\jtd~transition are required to check this
possibility.

\subsection{Turbulence in Outer Disks}

We derive intrinsic (local) line widths ranging between $0.12$ and $0.29 \kms$. When taking into account the
thermal component ($0.08$ to $0.15 \kms$), from Eqs.\ref{eq:turb2} and \ref{eq:turb3} we derive turbulent
widths below 0.15 $\kms$. These values should be used as upper limits, since the spectral resolution used for
the analysis (0.2 $\kms$) is comparable to the derived line widths. They are nevertheless significantly
smaller than the sound speed, $C_s = 0.3$ to $0.5 \kms$ in the relevant temperature and radius range. The
turbulence is thus largely subsonic. A more precise analysis, using the full spectral resolution and an
accurate knowledge of the kinetic temperature distribution, is required for a better determination.

\label{sec:turb}

\subsection{Hydrostatic Scale Height}

We follow the approach which is described in Sect.\ref{sec:scale} where the scale-height is not formally
derived from the hydrostatic equilibrium. In a first step, the flaring exponent $e_h$ is taken equal to
$e_h=-1.25$. Such a value is in agreement with predictions from models such as \citet{Chiang_etal1997} and
\citet{DAlessio_etal1999}. The observed transitions indicate apparent scale heights at 100 AU of 30 AU in
DM\,Tau, and 19 AU in LkCa\,15 and MWC\,480 (see Tab.\ref{tab:dmtau}-\ref{tab:mwc480}). The only discrepant
result is the value derived from \dco\ in MWC\,480, 10 AU. In a second step, treating $e_h$ as a free
parameter provides similar results, except for LkCa\,15 where the \tco\ data is best fitted by a flat
($e_h=0$), but thick disk (see Fig.\ref{fig:radlk}).

On the other end, using the temperature derived from the \tco~\jdu~ transition as appropriate, we can derive
the hydrostatic scales heights. These scale heights are 15, 13 and 9.5 AU, respectively for DM\,Tau, LkCa\,15
and MWC\,480, with a typical exponent $e_h=-1.35$, approximately 1.5 -- 2 times smaller than the apparent
values. Although, as mentioned in Sect.\ref{sec:scale}, the scale height has to be interpreted with some
caution (especially when determined from \dco), these results suggest that CO has a broader vertical
distribution than what is expected from an hydrostatic distribution of the gas. The current data do not allow
us to identify the causes of this effect. Vertical spreading due to turbulence can be rejected as the
turbulence is largely subsonic (see Sect.\ref{sec:turb}). The larger apparent height may be due to CO being
more abundant in the photo-dissociation layer above the disk plane.

\subsection{Comparison with other Disks and Models}
\label{sec:compare}

Only one other source has been studied at the same level in the CO isotopologues: AB\,Aur by
\citet{Pietu_etal2005}. The temperature behavior of the AB\,Aur disk is similar to that of the MWC\,480 disk,
although the AB\,Aur disk is warmer. Temperature derived from CO for all sources are summarized in
Fig.\ref{fig:temp}. As expected, the disk temperatures (midplane and surface) clearly increase with the
stellar luminosity. The A0/A1 star HD34282, mapped by \citet{Pietu_etal2003} in \dco~ also exhibits a hotter
disk surface while the TTauri star GM Aur \citep{Dutrey_etal1998} has a colder disk. From Table 1 of
\citet{Pietu_etal2005}, the \tco\ surface density in AB\,Aur at 300 AU is about $2\,10^{16}$ cm$^{-2}$, 4 to
8 times larger than in the DM\,Tau, MWC\,480 or LkCa\,15 disks. This larger surface density probably results
from a lack of depletion onto dust grains, since the temperature in the AB\,Aur disk remains well above the
CO condensation temperature at least up to 700 AU in radius, contrary to those of DM\,Tau, LkCa\,15 and
MWC\,480 which are much colder beyond 200 - 300 AU.

LkCa\,15 differs slightly from the general picture: no clear vertical temperature gradient is observed, and
the \tco\ surface density slope is much less steep (exponent $p = 1.5$) and the CO isotopologues indicate a
discrepant scale height behaviour. This may be linked to a peculiar disk geometry: \citet{Pietu_etal2006}
have reported a 50 AU cavity in the continuum emission from LkCa\,15, perhaps due to planetary formation or a
low mass companion. This peculiar geometry may affect the temperature structure of the surrounding disk, by
for example shielding it from stellar radiation behind a warm, thick, inner rim. Any definite conclusion
about the peculiarities of LkCa\,15 seems premature at this point.

Molecular abundances in proto-planetary disks have been studied by several authors. Disks are believed to be
dominated by a warm molecular layer, exposed to the UV radiation emanating from the central star.
\citet{Zadelhoff_etal2003} showed 2-D radiative transfer effects result in a stronger penetration of the UV
field into the disk than simple 1+1D models assumed.
A more refined model was published by \citet{Aikawa_Nomura_2006}, which use a time dependent chemistry taking
into account the thermal structure of the disk, 2-D radiative transfer and sticking onto grains. Their model
parameters are more appropriate for DM\,Tau. Despite its comprehensiveness, this model fails to represent the
observed surface densities: it predicts relatively shallow ($p \simeq 1$) radial distribution of the surface
density, in sharp contrast with the steep slope observed. However, their model only extends out to 300 AU,
while the constraint we obtain on the surface densities in DM\,Tau refer mostly to larger radii.

\citet{Semenov_etal2006} point out the importance of vertical mixing in explaining the molecular surface
densities. A similar study was conducted by \citet{Willacy_etal2006} who conclude that a diffusion
coefficient of order 10$^{18}$ cm$^2$s$^{-1}$ is required to bring the CO surface density in reasonable
agreement with observations. Vertical mixing may be essential in explaining why CO is detected at temperature
well below the freeze out value, and also why, despite very different temperature profiles, DM\,Tau and
MWC\,480 display similar CO surface densities.

None of these model studied the \tco\  isotopologue. \citet{HilyBlant_etal2006} used a simpler PDR model to
study the effects of grain size and stellar UV flux on the CO isotopologue abundances. In this static model,
self-shielding of H$_2$, CO and \tco\ is treated explicitly, but CO sticking onto grains is not included.
This model shows the importance of fractionation and reproduces better the observed surface densities.  In
particular, from their Fig.6, [\dco]/[\tco] ratio of order 10 -- 20 is obtained for large grains ($a_+ =
3\,10^{-4}$~m) and $\chi = 1000$ ($\chi$ being the enhancement over the interstellar UV field at 100 AU, with
a 1/r$^2$ dependence), i.e. when the UV field penetration is important.

So far, no published model reproduces the smaller outer radii for \tco\ than for \dco . To do so, it seems
essential to account for selective photodissociation of CO isotopologues by the interstellar UV radiation
field, and in particular \textit{by the UV radiation impinging on the disk outer edge}. The disappearance of
\tco\ at distances $\sim 500$ AU from MWC\,480 and LkCa\,15 indicate that the typical extinction through the
disk at this radius corresponds roughly to $A_V \simeq 1$, although the disks extend at least to 700 - 900
AU. A consistency check can be done on this hypothesis: assuming a standard $A_V/$H$_2$ ratio, the \tco\
outer radius can be used to constrain the CO abundance at this distance. From
Fig.\ref{fig:raddm}-\ref{fig:radmw}, the typical \dco\ surface density at the \tco\ outer radius is
$2-4\,10^{15}$~cm$^{-2}$, which converts to X[\dco] $\simeq 3-6\,10^{-5}$, in good agreement with the
expectation in this transition region. In DM\,Tau, $A_V \simeq 1$ would occur near 800 AU, i.e. the (outer)
DM\,Tau disk is more opaque than those of MWC\,480 and LkCa\,15. This larger opacity can result either from a
larger H$_2$ density, or from a larger $A_V/$H$_2$, i.e. more small grains in DM\,Tau.

All the chemical models are hampered by the need to specify the appropriate H$_2$ density distribution and
the dust properties. These parameters remain largely unknown.  The values of $p$ and $R_\mathrm{out}$ for the
dust in LkCa\,15 and MWC\,480 derived by \citet{Pietu_etal2006} differ systematically from those found here
for CO. In addition, the absolute value of the absorption coefficient of dust is also uncertain, which makes
the derivation of an H$_2$ surface density through continuum observations quite difficult. Using the
parameters given by \citet{Pietu_etal2006}, we find at the dust core outer radius, $\approx 180$~AU, that the
CO abundance is $\sim 10^{-6}$ for LkCa\,15, while it is $\sim3\,10^{-7}$--\,$3\,10^{-6}$ for MWC\,480,
depending on which temperature profile is adopted for the dust. These very low values, as well as the very
different values of the exponent $p$ for dust and CO, may thus indicate that the H$_2$ surface density cannot
be extrapolated by a simple power law, but displays instead a very significant change near $\sim 150$~AU.
This can have significant impact on the chemical model predictions for any specific disk.

\section{Summary}

We report new \tco~ observations in LkCa\,15 and MWC\,480 obtained with the IRAM array, and \hco~
observations of LkCa\,15 and MWC\,480 and DM\,Tau. These data were analyzed together with existing \dco~
data. As we improved our modelling method, we also carried again the analysis on DM\,Tau data as a
consistency check. By comparing these results with previous one obtained on AB\,Aur (and to the temperatures
derived from \dco{} observations in HD34282 and GM Aur), this allowed us to derive the following conclusions:

\begin{itemize}

\item We systematically detect radial gas temperature gradients. Vertical temperature gradients are also
ubiquitous in AB\,Aur, DM\,Tau, and MWC\,480, but LkCa\,15 is an apparent exception (perhaps due to its
peculiar geometry). Disks surrounding hotter stars, such as Herbig Ae objects (MWC\,480 and AB\,Aur), present
a hotter CO surface. This behavior agrees with the prediction from models of passive irradiated disks.

\item CO disks around hotter stars are also hotter around the mid-plane (AB\,Aur and to a lesser extent
MWC\,480). Although [CO/H$_2$] ratios are not directly measurable, the much higher surface density of \tco\
in the warmest disk, AB\,Aur, indicates that depletion onto grains is an important mechanism for the colder
disks. However, the CO surface density does not appear to be a monotonous function of the stellar type.

\item For the more sensitive transitions, we are able to measure  an apparent scale height which appears to
be larger than the hydrostatic scale height. Although marginal, this result is in agreement with the
expectation that molecules preferentially form in a warm photo-dissociation layer above the disk plane.

\item However, in the outer part of TTauri disks ($R \geq 300$~AU) and even around MWC\,480, the averaged
kinetic temperature measured along the line of sight is well below the freeze-out temperature of CO, but a
significant amount of CO is still observed. This is in contradiction with most predictions of chemical models
but would be in favor of models with vertical mixing.

\item The CO surface density falls off much more rapidly with radius than predicted by most chemical models.
In the cases of DM\,Tau and LkCa\,15, we also observe that the \dco~ has a steeper surface density exponent
than \tco . This suggests a steepening of the surface density slope beyond 300 -- 500 AU, but
photo-dissociation, and also perhaps depletion on grains are likely to play a role.

\item A change in surface density behaviour is also suggested by the very different parameters (exponent $p$
and outer radius) derived from the continuum emission.

\item For all disks where \dco~and \tco\ data are available, photo-dissociation by the ambient UV field
impinging on the disk edge is likely the appropriate mechanism to explain the smaller outer radii determined
in \tco\ compared to that of \dco.

\item Except for the warmest disks, CO fractionation is an important mechanism. For DM\,Tau and LkCa\,15, we
estimate an isotopologue ratio, of order or less than 10 -- 20 between 300 AU and the \tco\ outer radius.

\item At 300 AU, the [\tco]/[\hco] ratio is around 600, within a factor 2. It apparently decreases with
radius.

\end{itemize}

\begin{acknowledgements}
We acknowledge all the Plateau de Bure IRAM staff for their help during the observations. We also would like
to thank Guilen Oyar\c{c}abal and Rowan Smith for their participation to the fits.
\end{acknowledgements}

\bibliography{ms6537}

\begin{thebibliography}{36}
\expandafter\ifx\csname natexlab\endcsname\relax\def\natexlab#1{#1}\fi

\bibitem[{{Aikawa} \& {Nomura}(2006)}]{Aikawa_Nomura_2006}
{Aikawa}, Y. \& {Nomura}, H. 2006, \apj, 642, 1152

\bibitem[{{Asplund} {et~al.}(2004){Asplund}, {Grevesse}, {Sauval}, {Allende
  Prieto}, \& {Kiselman}}]{Asplund_etal2004}
{Asplund}, M., {Grevesse}, N., {Sauval}, A.~J., {Allende Prieto}, C., \&
  {Kiselman}, D. 2004, \aap, 417, 751

\bibitem[{{Chiang} \& {Goldreich}(1997)}]{Chiang_etal1997}
{Chiang}, E.~I. \& {Goldreich}, P. 1997, \apj, 490, 368

\bibitem[{{D'Alessio} {et~al.}(1999){D'Alessio}, {Calvet}, {Hartmann},
  {Lizano}, \& {Cant{\'o}}}]{DAlessio_etal1999}
{D'Alessio}, P., {Calvet}, N., {Hartmann}, L., {Lizano}, S., \& {Cant{\'o}}, J.
  1999, \apj, 527, 893

\bibitem[{{Dartois} {et~al.}(2003){Dartois}, {Dutrey}, \&
  {Guilloteau}}]{Dartois_etal2003}
{Dartois}, E., {Dutrey}, A., \& {Guilloteau}, S. 2003, \aap, 399, 773

\bibitem[{{Dullemond} \& {Dominik}(2004)}]{Dullemond_etal2004}
{Dullemond}, C.~P. \& {Dominik}, C. 2004, \aap, 417, 159

\bibitem[{{Dullemond} {et~al.}(2001){Dullemond}, {Dominik}, \&
  {Natta}}]{Dullemond_etal2001}
{Dullemond}, C.~P., {Dominik}, C., \& {Natta}, A. 2001, \apj, 560, 957

\bibitem[{{Dutrey} {et~al.}(1997){Dutrey}, {Guilloteau}, \&
  {Guelin}}]{Dutrey_etal1997}
{Dutrey}, A., {Guilloteau}, S., \& {Guelin}, M. 1997, \aap, 317, L55

\bibitem[{{Dutrey} {et~al.}(1998){Dutrey}, {Guilloteau}, {Prato}, {Simon},
  {Duvert}, {Schuster}, \& {Menard}}]{Dutrey_etal1998}
{Dutrey}, A., {Guilloteau}, S., {Prato}, L., {et~al.} 1998, \aap, 338, L63

\bibitem[{{Dutrey} {et~al.}(1994){Dutrey}, {Guilloteau}, \&
  {Simon}}]{Dutrey_etal1994}
{Dutrey}, A., {Guilloteau}, S., \& {Simon}, M. 1994, \aap, 286, 149

\bibitem[{{Duvert} {et~al.}(2000){Duvert}, {Guilloteau}, {M{\'e}nard}, {Simon},
  \& {Dutrey}}]{Duvert_etal2000}
{Duvert}, G., {Guilloteau}, S., {M{\'e}nard}, F., {Simon}, M., \& {Dutrey}, A.
  2000, \aap, 355, 165

\bibitem[{{Grady} {et~al.}(1999){Grady}, {Woodgate}, {Bruhweiler}, {Boggess},
  {Plait}, {Lindler}, {Clampin}, \& {Kalas}}]{Grady_etal1999}
{Grady}, C.~A., {Woodgate}, B., {Bruhweiler}, F.~C., {et~al.} 1999, \apjl, 523,
  L151

\bibitem[{{Grevesse} {et~al.}(2005){Grevesse}, {Asplund}, \&
  {Sauval}}]{Grevesse_etal2005}
{Grevesse}, N., {Asplund}, M., \& {Sauval}, A.~J. 2005, in EAS Publications
  Series, ed. G.~{Alecian}, O.~{Richard}, \& S.~{Vauclair}, 21--30

\bibitem[{{Guilloteau} \& {Dutrey}(1994)}]{Guilloteau_Dutrey1994}
{Guilloteau}, S. \& {Dutrey}, A. 1994, \aap, 291, L23

\bibitem[{{Guilloteau} \& {Dutrey}(1998)}]{Guilloteau_Dutrey1998}
{Guilloteau}, S. \& {Dutrey}, A. 1998, \aap, 339, 467

\bibitem[{{Guilloteau} {et~al.}(1999){Guilloteau}, {Dutrey}, \&
  {Simon}}]{Guilloteau_etal1999}
{Guilloteau}, S., {Dutrey}, A., \& {Simon}, M. 1999, \aap, 348, 570

\bibitem[{{Hartmann} {et~al.}(1998){Hartmann}, {Calvet}, {Gullbring}, \&
  {D'Alessio}}]{Hartmann_etal1998}
{Hartmann}, L., {Calvet}, N., {Gullbring}, E., \& {D'Alessio}, P. 1998, \apj,
  495, 385

\bibitem[{{Hily-Blant} {et~al.}(2007){Hily-Blant}, {Dartois}, {Roueff}, {Pineau
  des For\^ets}, {Dutrey}, \& {Guilloteau}}]{HilyBlant_etal2006}
{Hily-Blant}, P., {Dartois}, E., {Roueff}, E., {et~al.} 2007, \aap, submitted

\bibitem[{{Koerner} {et~al.}(1993){Koerner}, {Sargent}, \&
  {Beckwith}}]{Koerner_etal1993}
{Koerner}, D.~W., {Sargent}, A.~I., \& {Beckwith}, S.~V.~W. 1993, Icarus, 106,
  2

\bibitem[{{Malbet} {et~al.}(2001){Malbet}, {Lachaume}, \&
  {Monin}}]{Malbet_etal2001}
{Malbet}, F., {Lachaume}, R., \& {Monin}, J.-L. 2001, \aap, 379, 515

\bibitem[{{Mannings} {et~al.}(1997){Mannings}, {Koerner}, \&
  {Sargent}}]{Mannings_etal1997}
{Mannings}, V., {Koerner}, D.~W., \& {Sargent}, A.~I. 1997, \nat, 388, 555

\bibitem[{{Monnier} \& {Millan-Gabet}(2002)}]{Monnier_Millan2002}
{Monnier}, J.~D. \& {Millan-Gabet}, R. 2002, \apj, 579, 694

\bibitem[{{Pavlyuchenkov} {et~al.}(2007){Pavlyuchenkov}, {Guilloteau},
  {Henning}, {Pi\'etu}, {Semenov}, \& {Dutrey}}]{Pavlyuchenkov2007}
{Pavlyuchenkov}, Y., {Guilloteau}, S., {Henning}, T., {et~al.} 2007, in prep.

\bibitem[{{Pety}(2005)}]{pety05}
{Pety}, J. 2005, in SF2A-2005: Semaine de l'Astrophysique Francaise, ed.
  F.~{Casoli}, T.~{Contini}, J.~M. {Hameury}, \& L.~{Pagani}, 721--+

\bibitem[{{Pety} {et~al.}(2006){Pety}, {Gueth}, {Guilloteau}, \&
  {Dutrey}}]{Pety_etal2006}
{Pety}, J., {Gueth}, F., {Guilloteau}, S., \& {Dutrey}, A. 2006, \aap, 458, 841

\bibitem[{{Pi{\' e}tu} {et~al.}(2003){Pi{\' e}tu}, {Dutrey}, \&
  {Kahane}}]{Pietu_etal2003}
{Pi{\' e}tu}, V., {Dutrey}, A., \& {Kahane}, C. 2003, \aap, 398, 565

\bibitem[{{Pi{\'e}tu} {et~al.}(2006){Pi{\'e}tu}, {Dutrey}, {Guilloteau},
  {Chapillon}, \& {Pety}}]{Pietu_etal2006}
{Pi{\'e}tu}, V., {Dutrey}, A., {Guilloteau}, S., {Chapillon}, E., \& {Pety}, J.
  2006, \aap, 460, L43

\bibitem[{{Pi{\'e}tu} {et~al.}(2005){Pi{\'e}tu}, {Guilloteau}, \&
  {Dutrey}}]{Pietu_etal2005}
{Pi{\'e}tu}, V., {Guilloteau}, S., \& {Dutrey}, A. 2005, \aap, 443, 945

\bibitem[{{Qi} {et~al.}(2003){Qi}, {Kessler}, {Koerner}, {Sargent}, \&
  {Blake}}]{Qi_etal2003}
{Qi}, C., {Kessler}, J.~E., {Koerner}, D.~W., {Sargent}, A.~I., \& {Blake},
  G.~A. 2003, \apj, 597, 986

\bibitem[{{Semenov} {et~al.}(2005){Semenov}, {Pavlyuchenkov}, {Schreyer},
  {Henning}, {Dullemond}, \& {Bacmann}}]{Semenov_etal2005}
{Semenov}, D., {Pavlyuchenkov}, Y., {Schreyer}, K., {et~al.} 2005, \apj, 621,
  853

\bibitem[{{Semenov} {et~al.}(2006){Semenov}, {Wiebe}, \&
  {Henning}}]{Semenov_etal2006}
{Semenov}, D., {Wiebe}, D., \& {Henning}, T. 2006, \apjl, 647, L57

\bibitem[{{Siess} {et~al.}(2000){Siess}, {Dufour}, \&
  {Forestini}}]{Siess_etal2000}
{Siess}, L., {Dufour}, E., \& {Forestini}, M. 2000, \aap, 358, 593

\bibitem[{{Simon} {et~al.}(2000){Simon}, {Dutrey}, \&
  {Guilloteau}}]{Simon_etal2000}
{Simon}, M., {Dutrey}, A., \& {Guilloteau}, S. 2000, \apj, 545, 1034

\bibitem[{{van den Ancker} {et~al.}(1998){van den Ancker}, {de Winter}, \&
  {Tjin A Djie}}]{VandenAncker_etal1998}
{van den Ancker}, M.~E., {de Winter}, D., \& {Tjin A Djie}, H.~R.~E. 1998,
  \aap, 330, 145

\bibitem[{{van Zadelhoff} {et~al.}(2003){van Zadelhoff}, {Aikawa},
  {Hogerheijde}, \& {van Dishoeck}}]{Zadelhoff_etal2003}
{van Zadelhoff}, G.-J., {Aikawa}, Y., {Hogerheijde}, M.~R., \& {van Dishoeck},
  E.~F. 2003, \aap, 397, 789

\bibitem[{{Willacy} {et~al.}(2006){Willacy}, {Langer}, {Allen}, \&
  {Bryden}}]{Willacy_etal2006}
{Willacy}, K., {Langer}, W., {Allen}, M., \& {Bryden}, G. 2006, \apj, 644, 1202

\end{thebibliography}
\bibliographystyle{aa}

\end{document}